\documentclass[aps,pra,amssymb,twocolumn,showpacs,a4paper]{revtex4-1}

\usepackage{amsmath,amssymb,epsfig,graphicx,color,framed}
\usepackage[english]{babel}
\usepackage{hyperref}
\usepackage{amsfonts} 
\usepackage{amsmath}
\usepackage{amssymb}
\usepackage{graphicx}
\usepackage{hyperref}
\usepackage{color}
\usepackage{natbib}  
\usepackage{bm} 
\usepackage{isomath} 
\usepackage{braket}
\usepackage{tikz-cd}
\usepackage{amsmath,amscd}
\usepackage{adjustbox}

\newcommand{\vect}[1]{\vectorsym{#1}} 

\begin{document}

\title{Topological magnetoelectric response in passive magnetic devices}

\author{Antonio A. Valido}
\email{alejandro.valido@urjc.es}
\affiliation{Nonlinear Dynamics, Chaos and Complex Systems Group,
Departamento de F\'isica, Universidad Rey Juan Carlos,
Tulip\'an s/n, 28933 M\'ostoles, Madrid, Spain}

\author{Alejandro J. Castro}
\email{alejandro.castilla@nu.edu.kz}
\affiliation{Department of  Mathematics, Nazarbayev University,
		010000 Astana, Kazakhstan}

\date{\today}

\keywords{}
\pacs{}

\begin{abstract}
Despite the prospect of next-generation electronic technologies has spurred the investigation of the remarkable topological magnetoelectric response, it remains largely unexplored its potential in the application of basic electronic devices. In this paper, we undertake this task at the theoretical level by addressing the $\theta$-electrodynamics and examine electromagnetic properties (e.g. tunable inductance, operating frequency range, and power consumption) of three fundamental passive magnetic devices endowed with this effect: the primitive transformer, the bilayer solenoid inductor, and the solenoid actuator. We further exploit the methodology of magnetic circuits to obtain an extended Hopkinson's law that is valid for both topological and ordinary magnetoelectric responses (provided it is uniform in the bulk). Under low-power conditions, we find out that the functionally passive part of the topological-magnetoelectric transformer, solenoid inductor as well as solenoid actuator is indistinguishable from the conventional situation up to second-order in the magnetoelectric susceptibility; and argue that the main
benefit of using topological insulators essentially relies on a lower power consumption. Our theoretical framework is also convenient to analyse magnetoelectric inductors endowed with a relatively large magnetoelectric susceptibility,  they display a broad inductance tunability of over $200\%$ up to $100$ GHz in the millimeter length scale. Conversely, our treatment predicts that the operating frequency range could be restricted below the ultra low frequency by a significantly strong magnetoelectric response (e.g. retrieved by certain multiferroic heterostructures). 
\end{abstract}

\maketitle

\section{Introduction}

During the last decades theoretical and experimental efforts have been devoted to envisage new materials enable to exhibit a (linear) magnetoelectric (ME) response \cite{ying20221,liang20211,tokura20071,yan20181}, where an electric field is induced by an applied magnetic field (and vice versa) in a way that cannot be explained by the standard Maxwell's equations \cite{fiebig20051}. To date such response has been experimentally observed in a variety of materials that break time-reversal and inversion symmetries \cite{fiebig20051,ying20221}: from the originally suggested antiferromagnetic model Cr$_{2}$O$_3$ \cite{dzyaloshinskii19601} to recently introduced multiferroic compounds such as FeRh/BaTiO$_3$ \cite{cherifi20141}. Remarkably, the ME effect has been theoretically proposed to be realized in a fascinating class of electronic materials called topological insulators (TIs). These are characterized by having a robust insulating bulk and hosting non-dissipative surface charge and spin currents \cite{qi2010}. Concretely, it was shown that time-reversal-symmetric TIs \cite{qi20081,essin20091} and axion insulators (AIs) in three dimensions (3D) feature a ME response which is uniform and quantized in terms of the fine-structure constant in the bulk (despite they preserve the aforementioned symmetries), this is called the topological magnetoelectric (TME) response. While the latter has not yet been directly observed since it requires careful experiments \cite{armitage20191}, it has measurable consequences (e.g. the quantized Faraday and Kerr rotation) that have been experimentally detected for stationary electromagnetic fields \cite{wu20161}.

Nowadays, it is widely accepted that the interface of various magnets with TIs will find applications in the development of next-generation quantum technologies \cite{moore20101,he20221,nogueira20161,nogueira20181} (e.g. quantum information and communication or topological spintronics). For instance, the combination of TIs with superconductors constitutes a promising platform to build the long-desired quantum computers in the long term \cite{pachos20121}.  On the other hand, non-topological ME materials are intensively investigated for the development of innovative applications in several areas, ranging from passive magnetic electronic to low-power spintronics \cite{fusil20141}. In particular, there is an increasing interest in integrated magnetic devices, such as tunable inductors \cite{zare20151,lou20091,chen20201,su20161,yan20201}, actuators, or transformers \cite{yang20211,fetisov20221,luo20211}: these provide a new paradigm for 
circuit design of adaptive power converters (e.g. the transformer) or tunable multiband radio-frequency (RF) communications systems \cite{chen20201,liang20211,liang20212,he20211}. In this context, a natural question arises as to what extend TME materials are also suitable candidates to manufacture passive electronic components or magnetic devices endowed with certain desirable properties \cite{gilbert20211,he20221,breunig20211,tian20171,tokura20191}: a high inductance, a high operating frequency (e.g. $1-10^{8}$ Hz) and a  low power consumption.  

\begin{figure}[t]
    \includegraphics[width=0.45\textwidth]{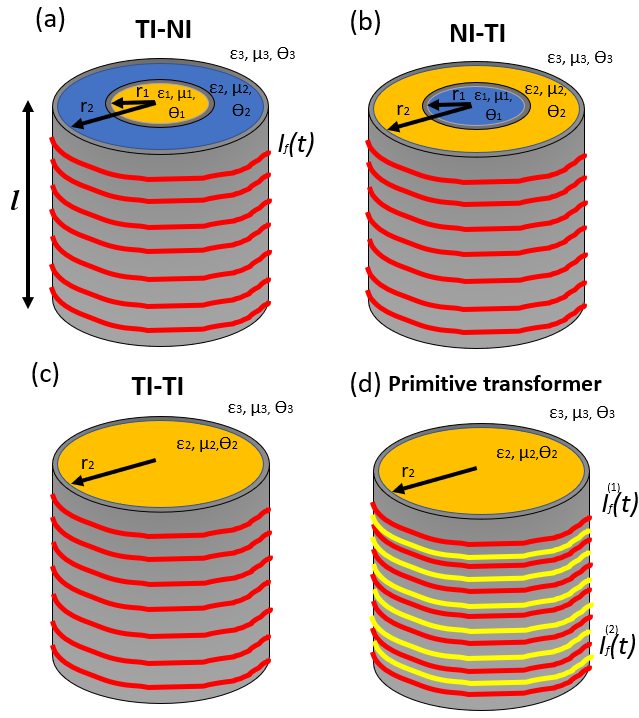}
  \caption{(color online) Illustration of the three bilayer solenoid inductors (see figures (a), (b) and (c)) and the primitive transformer (see figure (d)). The orange layer represents a ME material characterized by certain dielectric constant $\varepsilon_r(\omega)$, relative permeability $\mu_{r}$ and 
 ME polarizability $\theta$ (see discussion in Sec. \ref{SecTME}); whereas the ordinary insulator with dielectric contant $\tilde\varepsilon_{r}$ is indicated by the blue layer.  Notice that these arrangements resemblance tunable inductors consisting of ME heterostructures and ME laminates. The red lines represent the coil which is fed by an alternating current $I_{f}(t)$. In the primitive transformer, the red coil represents the primary winding, while the yellow lines correspond to the secondary winding. Although it is not shown here, we also study the primitive transformer endowed with the same three configurations of the core illustrated for the solenoid inductor. We shall consider, as usual, that the solenoid length $l$ is sufficiently large in comparison with the solenoid radio $r_{2}$ (not illustrated), so we may ignore the so-called fringe effects at the ends of the solenoids. \label{Fig1Setup}} 
\end{figure}

Motivated by this question, the present work is devoted to elucidate the potential of passive magnetic devices composed of ME materials, paying special attention to time-reversal-symmetric TIs or AI (e.g. 
 Bi$_{2}$Te$_{3}$, TlBiSe$_2$ or MnBi$_{2}$Te$_{4}$ compounds \cite{tokura20191,he20221}). For concreteness, we assess the induction coefficients, the (time-averaged) electromagnetic energy, and the (time-averaged) power radiated for the TME solenoid inductor; the induced electromotive force of the TME primitive transformer (sketched by Fig. (\ref{Fig1Setup}.d)); and the magnetic force strength of the TME solenoid actuator when they are excited by an alternating current (ac) source with a tunable frequency. By starting from the so-called $\theta$-electrodynamics \cite{martin20191,martin20182} representing a low-energy effective description of the (linear) TME response \cite{nenno20201,sekine20211,zirnstein20201,armitage20191}, we carry out an extensive numerical and perturbative analyses for strong and weak ME effects
as well as for large and small frequencies (e.g. the RF frequency range). Interestingly, we find out that the Ampère's law approximately holds by taking account a surface Hall current responsible for an uniform ME response (in the bulk). Relying on this result, we derive an extended Hopkinson's law (which represents the counterpart of the Ohm's law in magnetic circuits) applicable to a broad class of ME devices, including both topological and non-topological systems, in the RF frequency range. Overall, the tunability of the TME solenoid inductor is significantly poor compared with previous magnetocrystallines under low-power conditions \cite{yan20181}, and the functionally passive part of the TME transformer as well as TME solenoid actuator coincides with the conventional situation \cite{zahn19791,kidd20201} up to second-order in the ME susceptibility. We then argue that the central benefit of using TME materials is a lower power consumption as undesired eddy currents must disappear owing to the fact that the bulk conduction is suppressed in topological insulators by construction \cite{he20221}.  Conversely, we show that the generation of highly-intense electromagnetic fields becomes energetically expensive for significant ME responses such as in 2D multiferroic heterostructures FeRh/BaTiO$_3$, imposing a cutoff frequency in the very low frequency domain (i.e. $1-10^6$ Hz) for millimeter length scales. Let us stress that, though our motivation is in the spirit of previous works as Refs. \cite{philip20171,gilbert20211,medel20231}, we dealt with different systems and follow a drastically distinct approach (while they provide a microscopic treatment, we carry out a macroscopic description relying on the constitutive relations). We also notice that our results have none counterpart in previous treatments addressing the TME response in the magnetostatic situation \cite{martin20191,martin20181,martin20182,nogueira20221,ryu20121,zeng20101}, since the ME response here represents a pure dynamical effect (i.e. it cancels in the limit of direct current sources).

\section{$\theta$-electrodynamics in passive magnetic devices}

The most simple as well as general scheme of a magnetic devices is the primitive transformer sketched in Fig. (\ref{Fig1Setup}.d), this serves as an archetypal model of a magnetic circuit extensively addressed in many advanced textbooks as well as peer reviews \cite{rivera20021,jackson20121,batell20121,kidd20201,wagemakers20171}. Basically, this consists of two coils winding a permeable core enable to completely confine the axial magnetic field. This can be further decomposed  into two (infinitely-long) solenoid inductors that must encapsulate the essential features of the TME response displayed by the interesting magnetic circuit. For sake of readability, we shall employ the solenoid inductor as a testbed to explore the electromagnetic phenomena arising from the $\theta$-electrodynamics. Concretely, we consider the three setups sketched in Fig. (\ref{Fig1Setup}.a), (\ref{Fig1Setup}.b), and (\ref{Fig1Setup}.c). These consist of a long tall solenoid of length $l$ with a tightly wounded thin wire and surrounded by vacuum. Importantly, these can be composed of two cylindrical layers of radios $r_{i}$ ($i=1,2$) made by either trivial insulators or TIs featuring the TME response: more precisely,  the inner and outer layer correspond to the TI in the TI-NI (see Fig. (\ref{Fig1Setup}.a)) and NI-TI (see Fig. (\ref{Fig1Setup}.b)) setups, respectively; whereas the TI-TI system represents the scenario in which the solenoid bar is just built by the TI (see Fig. (\ref{Fig1Setup}.c)). In all the studied situations, the coil is fed by an alternating current, say $I_{f}(t)=I_{0}\cos(\omega_{0}t)$, characterized by certain frequency $\omega_{0}$. Let us remark that the solenoid inductor represents one of the basic passive components in electronic circuits and finds many applications in RF circuits \cite{he20211,zare20151,chen20201,harmon19911,lin20211,mcdonald19971,thevenet19991,su20161,yan20201} (e.g. in impedance matching working or low-noise amplifier).

We will now proceed in three steps. The following section will settle the basic formalism behind the $\theta$-electrodynamics paying special attention to the experimental conditions, and introduce the interesting electromagnetic magnitudes studied through the rest of the manuscript. The subsequent section \ref{Sec_MRMC} will contain a brief summary of the main results of the present work related to the magnetic circuits, e.g. the extended Hopkinson's law for ME devices, and provide the magnetic circuit diagram corresponding to the TME primitive transformer. Later in Sec. \ref{Sec_BLS}, we will solve the $\theta$-electrodynamics for the aforementioned solenoid inductors, and carried out an extensive numerical and perturbative analysis of the electromagnetic fields, induction coefficients, the electromagnetic energy, and the power radiated for high and low frequencies, as well as for strong and weak ME responses. Notice that this analysis will encompass the electromagnetic properties of both topological and non-topological ME materials. Sec. \ref{Sec_MCC} addresses the induced electromotive force in a ideal TME primitive transformer, the linkage magnetic flux of a real TME transformer and the magnetic force strength appearing in the TME solenoid actuator by making use of the preceding analysis, which leads to the main results previously presented in Sec. \ref{Sec_MRMC}. Finally, we draw the main conclusions and summarize the results in a table in Sec. \ref{SecCOUT}.

\subsection{Preliminaries}\label{SecTME}

The ME effect essentially consists of a coupling between the electric and magnetic fields which enables to induce an electric polarization when applying a magnetic field or vice-versa \cite{liang20211, martin20191,armitage20191}. This can be most readily seen in terms of the constitutive relations characterizing the (linear) electromagnetic response against weak and slowly time-varying electric $\vect E$ and magnetic $\vect B$ fields. Concretely, we shall focus in ME media whose (macroscopic) constitutive relations read \cite{ying20221,fiebig20051} 
\begin{eqnarray}
\vect D&=&\varepsilon_{r}\varepsilon_{0}\vect E+\chi\vect B, \label{ECT2} \\
\vect H&=&\frac{1}{\mu_{r}\mu_{0}}\vect B-\chi\vect E, \label{ECT1}
\end{eqnarray}
where $\varepsilon_{r}$ and $\mu_{r}$ are the relative electric permittivity (or dielectric constant) and the relative magnetic permeability of the medium, whereas $\varepsilon_{0}$ and $\mu_{0}$ are the familiar permittivity and permeability of the vacuum. Here, $\chi$ denotes the (linear) ME susceptibility \cite{fiebig20051}. It is convenient to express the latter in terms of the fine-structure constant, such that $\chi=\frac{\theta}{\pi}\alpha_{0}$ with $\alpha_{0}=\frac{e^2}{4\pi\varepsilon_{0}\hbar c}\sqrt{\frac{\varepsilon_{0}}{\mu_{0}}}$ (where $e$, $\hbar$ and $c$ denote the elementary charge, the Planck's constant and the vacuum light speed, respectively) and $\theta$ being a dimensionless parameter distinctive of the ME medium which is called ME polarizability or axion angle \cite{martin20191} (this can be though of a characteristic of the material similarly as $\mu_r$ and $\varepsilon_r$). The TME effect is refereed to the case when $\theta$ takes a constant quantized value $\theta=\pi(\text{mod} \ 2\pi)$ in the bulk \cite{bhattacharyya20211,sekine20211,zirnstein20201} (while it is $\theta=0(\text{mod} \ 2\pi)$ in a vacuum or a topologically trivial insulator). Importantly, time-reversal invariant TIs (e.g. the paradigmatic model Bi$_{2}$Se$_{3}$) and AIs (e.g. the antiferromagnetic compound MnBi$_2$Te$_4$) have been predicted to display the TME response below the Néel temperature of the material (e.g. it is theoretically estimated to be around $25$ K for MnBi$_2$Te$_4$ \cite{tokura20191,sekine20211}).  Recent experimental and theoretical progresses have also shown that the ordinary ME response (i.e. $\theta$ is no longer quantized) manifests in a multitude of both topological and non-topological materials in two and three dimensions \cite{ying20221}: for instance, some magnetic TIs \cite{tokura20191} or the previously mentioned antiferromagnetic model Cr$_{2}$O$_3$ (which exhibits a ME coupling coefficient $\theta\sim \frac{\pi}{36}$ \cite{wu20161,sekine20211}). In particular, the quest for a gigantic magnetoelectric polarizability at room temperature has boosted the investigation on the design of multiferroic materials (combining ferroelectric and ferromagnetic parts) \cite{spaldin20191,fusil20141}, where it has been reached values of the ME polarizability $\theta\sim 2.1 \times 10^{6}$ or, equivalently, $\chi\sim 6.6 \times 10^{5}\alpha_{0}$  \cite{bhattacharyya20211,ying20221} (e.g. in the heterostructure multiferroic FeRh/BaTiO$_3$ \cite{cherifi20141}). Although the latter have established the strongest ME response experimentally achievable, some preliminaries works envisage novel topological heterostructures, which are based on wrapped 3D quantum anomalous Hall insulators, enable to achieve larger ME coupling strengths than $\theta\sim \pi$ \cite{nenno20201}. 

Additionally, though the dielectric loss between the kHz and THz ranges is negligible at sufficient low temperatures in certain ME materials (for instance, below 175K in the multiferroics BiFeO$_3$ \cite{kamba20071} or DyMnO$_3$ \cite{kagawa20091}), we shall take account weak dispersive effects as they can be specially important in the description of electromagnetic shielding in certain TI \cite{medel20231} (see also \cite{enriquez20221}). Frequency dependent dielectric analysis of various ME materials showed dispersion in the low frequency region that is well described by the celebrated Debye-Lorentz model \cite{ortega20091,li20061}, which phenomenologically explains the energy dissipation (i.e. absorption) of certain resonant frequencies due to ohmic losses. According to the latter, the dielectric constant dependence with the frequency in most practical situations exhibits a monodispersive relation  \cite{jackson20121,zangwill20121}
\begin{equation}
    \varepsilon_{r}(\omega)=1+\frac{\omega_{e}^2}{\omega_{R}^2-\omega(\omega+i\gamma_{0})},
    \label{EDCME}
\end{equation}
where $\gamma_{0}$ is the damping rate of the resonance at frequency $\omega_{R}$, and $\omega_{e}$ stands for the frequency strength \cite{medel20231} (which is also refereed to as the plasma frequency of the dielectric medium \cite{jackson20121,zangwill20121}). Interestingly, it turns out that strong absorption effects are negligible for several ME materials in the frequency domain under consideration (i.e. $\gamma_{0}\ll\omega_{R}$) \cite{fiebig20051,medel20231,liang20211}, so that  Eq. (\ref{EDCME}) can be thus approximated to
\begin{equation}
\varepsilon_{r}(\omega)\approx\frac{\omega_{R}^2+\omega_{e}^2-\omega^2}{\omega_{R}^2-\omega^2},
\label{EQDISRLF}
\end{equation}
which clearly becomes singular at the resonant frequency $\omega_{R}$. Similarly, the dielectric constant at frequencies far above the latter (i.e. $\omega\gg\omega_{R}$) reduces to the simple form \cite{jackson20121}
\begin{equation}
\varepsilon_{r}(\omega)\approx 1-\Big(\frac{\omega_{e}}{\omega}\Big)^2.
\label{EQDISRHF}
\end{equation}
We shall employ expressions (\ref{EQDISRLF}) and (\ref{EQDISRHF}) to study the ME response below as well as above the resonant frequency $\omega_{R}$, respectively. In particular, for numerical computation purposes we shall consider ME materials for which $\omega_{R}$ is embedded in the THz range; for instance, the topological insulator TlBiSe$_2$ for which $\omega_{R}\sim 1.6$ THz and $\omega_{e}=\sqrt{3}\omega_{R}$ \cite{medel20231}, or the conventional magnetoelectric BiFeO$_3$ for which $\omega_{R}\sim 1.3$ THz and $\omega_{e}\sim \sqrt{25}\omega_{R}$ \cite{kamba20071} (see further references in \cite{dong20151}). In other words, we will focus the attention on the frequency domain away from the resonant frequencies of the ME medium, so that the dispersive effects of the ME medium are sufficiently weak. Let us emphasize that this prescription is consistent with practical purpose since, for instance, most applications of the conventional ME effect occurs in the microwave range of frequencies \cite{fiebig20051}.

The experimental realization of the TME effect requires that both the TI surface breaks time-reversal symmetry and the surface Dirac states is maintained gapped \cite{sekine20211,nenno20201}. These can be achieved by using commensurate out- and in- plane antifferomagnetic or ferrimagnetic thin films, see Ref. \cite{oroszlany20121}.  In other words, the TME effect is observable as long as $\theta$ varies spatially in our setup (or temporally) \cite{zirnstein20201}. Under these working conditions, the topological (near-quantized) ME effect in the bulk stems from a surface response after applying either external magnetic or electric fields. More precisely, the TME effect relies on a non-dissipative surface Hall current, whose density reads \cite{essin20091,sekine20211,nenno20201,schultz20211}
\begin{equation}
    \vect J_{\text{Hall}}=\frac{\alpha_{0}}{\pi}\nabla \theta\times\vect E, \label{SHC}
\end{equation}
which emerges in the interfaces between the topological trivial and non-trivial insulators (i.e. this is the axion-induced effective current). To see this we must realize that, on one hand, the surface Hall current can be though of as a magnetization bound current density $\vect J_{b}$ by paying attention to the second term on the right-hand side of (\ref{ECT1}) (i.e. $\vect J_{b}=\frac{\alpha_0}{\pi}\nabla\times(\theta\vect E)$), and on the other hand, $\nabla \theta$ give rise to a spatial Dirac delta function since $\theta$ is considered to be a piece-wise constant function (see Eq. \ref{AAG}).  From this point onward we assume that the aforementioned experimental conditions are guaranteed so that TME effects are detectable in the interesting setups.

Importantly, we shall focus our attention on the frequency range in which the frequency dependence of the ME susceptibility is negligible. For instance, this is the THz regime for TI manifesting a quantized ME response (i.e. this occurs for TI with surface band gaps of the order of $10-100$ meV \cite{qi20111,crosse20171}). Additionally, it has been argued that the constitutive relations (\ref{ECT1}) and (\ref{ECT2}) characterize the electromagnetic response of experimentally feasible TIs at certain $\omega_{0}$ \cite{armitage20191}. On the other hand, it has been experimentally tested that the frequency dependence of the ME susceptibility is moderate within $10^1-10^4$ Hz for a broad class of conventional ME materials \cite{fiebig20051}, so that we can assume the conventional ME response (\ref{ECT1}) and (\ref{ECT2}) is independent of $\omega_{0}$. Beside avoiding strong absorption effects, we also restrict our future study to the frequency domain where (\ref{ECT2}) and (\ref{ECT1}) largely hold: we shall consider $\omega_{0}$ lying in the THz domain for TI (i.e. frequency range spanning $0.1-24.1$ THz), whereas our results related to conventional ME materials apply when $\omega_{0}$ could be roughly correspond to the RF domain.

Alternatively, the electromagnetic response of both topological and non-topological ME materials characterized by the constitutive relations (\ref{ECT1}) to (\ref{ECT2}) can be equivalently described in terms of the $\theta$-electrodynamics \cite{wilczek19871,ouellet2019,sekine20211,qi20081,armitage20191,qi20131} (even though time-reversal symmetry is broken \cite{baasanjav20141}), that is (expressed in ISQ unit)
\begin{eqnarray}
\nabla \cdot( \varepsilon\vect E)&=&\rho_{f}-\frac{\alpha_{0}}{\pi}\vect B\cdot \nabla \theta, \label{ME1} \\
\nabla \cdot \vect B &=& 0, \label{ME2} \\
\nabla \times \vect E&=&-\frac{\partial \vect B}{\partial t}, \label{ME3} \\
\nabla \times \Big(\frac{\vect B}{\mu_{r}\mu_{0}}\Big)&=&\vect J_{f}+\varepsilon_{r}\varepsilon_{0}\frac{\partial \vect E}{\partial t}+\frac{\alpha_{0}}{\pi}\nabla \theta\times \vect E , \label{ME4}
\end{eqnarray}
where  $\rho_{f}$ is the free density charge, $\vect J_{f}$ is the free current density  (e.g. coming from an external electric power supply). Clearly, the Ampère-Maxwell equation is modified by the magnetization current appearing in  (\ref{ECT1}), as somehow expected. It is important to realize that this represents a bulk current if $\theta$ changes continuously in the material volume; otherwise, it corresponds to a surface Hall current flowing in the material boundaries (in particular, the latter is half-quantized in the case of time-reversal-invariant TIs and AIs as anticipated above). Unlike, the Faraday's law remains unmodified. For sake of simplicity, it is instructive to rewrite this in terms of the induced electromotive force, denoted by $\epsilon$, as follows
\begin{equation}
    \epsilon=L\frac{d I_{f}}{dt}+N_{\text{rad}}\frac{d^2 I_{f}}{dt^2},
    \label{Eqfemlow}
\end{equation}
where $L$ is the familiar self-induction coefficient, and $N_{\text{rad}}$ stems for the magnetic induction arising from the displacement current \cite{brainerd19341} (i.e. it vanishes in the strict magnetostatic regime). In spite of the additional terms in the modified Maxwell's equations, the electromagnetic energy $U$ is given by the conventional expression as well, i.e. 
\begin{equation}
    U=\frac{1}{2}\int_{\mathcal{V}}\bigg(\varepsilon_{0}\frac{d(\omega\varepsilon_{r}(\omega))}{d\omega}|\vect E|^2+\frac{|\vect B|^2}{\mu} \bigg)d^3r,
    \label{EUEM}
\end{equation}
where we have explicitly taken account the dispersion of the ME media \cite{enriquez20221}. Here, $\mathcal{V}$ represents the spatial region occupied by the medium and $\mathcal{S}$ its boundary. More specifically, by starting from the usual definition of the electric work \cite{zangwill20121} and making use of Eqs. from (\ref{ME1}) to (\ref{ME4}), we arrive to the well-known Poynting's theorem, i.e.
\begin{equation}
    \frac{d W_{mech}}{dt}+\frac{d U}{d t}=-\oint_\mathcal{S} \vect S\cdot d\vect s, \label{EEQPT}
\end{equation}
with $W_{mech}$ denoting the mechanical work,  whereas $\vect S$ is the usual Poynting vector,
\begin{eqnarray}
\vect S=\frac{1}{\mu_{0}}\vect E\times \vect B, \label{ESEM}
\end{eqnarray}
which represents the power radiated per unit length. Notice that the energy dissipation due to strong absorption effects can be neglected for the present purpose since it holds that $\gamma_{0}\ll \omega_{R}$ for most interesting situations, as stated before. Equation (\ref{EUEM}) coincides with the expression for the usual electromagnetic energy in $\theta$-electrodynamics obtained in \cite{nikitin20121,patkos20221,fedorov20081}. As an alternate route, Eq. (\ref{EEQPT}) can be derived from the Lagrangian density associated to the $\theta$-electrodynamics (see App. \ref{app1}). In particular, since we are dealing with ac electromagnetic fields, we shall focus the attention on the time average of the induced electromotive force, the electromagnetic energy and the power radiated over a period of the current frequency (see Eqs. (\ref{EUEMA}),(\ref{PWRC}) and (\ref{EEEMTF})).

\subsection{Brief summary of results}\label{Sec_MRMC}
Our main objective is to provide a magnetic circuit analysis of experimentally feasible magnetic devices composed of materials displaying the TME response and excited by a sufficiently slow time-varying current density, namely $\vect J_{f}(t)$, so that we can employ the magnetoquasistatic approximation. In the conventional electrodynamics, this basically consists of neglecting the displacement current (i.e. $\vect J_{disp}=\varepsilon_{r}\varepsilon_{0}\frac{\partial \vect E}{\partial t}$) in Eq.(\ref{ME4}) \cite{ouellet2019,zangwill20121}, so we eventually recover the well-known Ampère's law, i.e. $\nabla \times \vect B=\mu_{r}\mu_{0} \vect J_{f}$. As anticipated in the introduction, we find out that a modified version of the latter holds in presence of the ME response as well. Concretely, in Sec. \ref{SecLFR} it is extensively shown for the interesting solenoid inductors that the axial component of the magnetic field, denoted by $B_{z}$, satisfies the modified Ampère's law,
\begin{equation}
\frac{\partial}{\partial r} \bigg(\frac{ B_{z}}{\mu_{r}\mu_{0}}\bigg)=-( J_{f,\varphi}+J_{\text{Hall},\varphi}), \label{MEQS4}
\end{equation}
where $J_{f,\varphi}$ and $J_{\text{Hall},\varphi}$ represent the azimuthal components of the free and non-dissipative surface Hall currents, respectively. Eq. (\ref{MEQS4}) holds for sufficiently low frequencies and weak ME polarizabilities, i.e.
\begin{align}
    \frac{r_{2}\omega_{0}}{c_{i}}\ll & 1, \label{DEEC} \\
    r_{2}\mu_{0}\omega_{0}\chi\ll & 1,  \label{MEWC}
\end{align}
with $c_{i}$ (for $i=1,2,3$) being the light speed of the media (notice that $c_{3}=c$). More specifically, we show that Eq. (\ref{MEQS4}) is valid up to first-order in the exciting frequency and second-order in the ME coupling strength. In practice, expression (\ref{DEEC}) is fulfilled for millimeter as well as micrometer laboratory solenoids (i.e. $r_{2}\sim 10$ mm or $r_{2}\sim 10 \ \mu$m) with current frequencies less than $300$ GHz or $10$ THz \cite{harmon19911,zangwill20121}, respectively; whereas the inequality (\ref{MEWC}) can be interpreted as a condition on the highest frequency that retrieves a weak ME response. To understand the latter we must realize that the ME response is a dynamical effect in the interesting devices, such that it cancels in the strict magnetostatic limit (i.e. $\omega_{0}\rightarrow 0$). The condition (\ref{MEWC}) is satisfied by conventional and topological ME materials at different frequencies and length scales: for instance, recalling that  $\chi= \alpha_{0}$ for time-reversal-invariant TIs and AIs (e.g., $\chi\sim 1/137\sqrt{\frac{\varepsilon_{0}}{\mu_{0}}}$ and $\omega_{R}\sim 1,6$ THz in the TlBiSe$_{2}$), our perturbative analysis is limited to exciting frequencies $\omega_{0}$ and solenoid lengths $r_{2}$ below $1.6$ THz and $10 \ \mu$m, respectively. For nontopological ME media with higher values of the ME susceptibility (e.g. $\chi\sim 4.8 \times 10^{3}\sqrt{\frac{\varepsilon_{0}}{\mu_{0}}}$ in the heterostructure multiferroic FeRh/BaTiO$_3$ \cite{cherifi20141,ying20221}), our treatment holds below exciting frequencies of $300$ GHz and solenoids lengths of $ 10 \ \mu$m, which completely covers the frequency range of state-of-the-art power electronics \cite{yan20181}.

According to (\ref{MEQS4}), we recover the standard expression of the axial magnetic field inside the solenoid, but taking account the azimuthal surface Hall current (see Eqs. (\ref{BzMQAa}) and (\ref{BzMQAb}) in Sec. \ref{SecLFR}). Intuitively, this means that $J_{\text{Hall},\varphi}$ represents an additional source of the magnetic flux across solenoid section, or equivalently, a magnetomotive force. Notice that this result applies to a magnetic core of arbitrary relative permeability $\mu_{r}$ (such as AIs) and endowed with a coil-carrying current, say $I_{f}$. By assuming the usual prescriptions of (i) a very highly permeable core (i.e. $\mu_{r}\gg 1$) and (ii) by neglecting fringing field effects (so that $B_{z}$ remains almost uniform and mostly lies within the solenoid core), we can perform a magnetic circuit analysis in presence of the ME response. Following the standard procedure \cite{zahn19791,kidd20201,yan20201},  Eq. (\ref{MEQS4}) leads to an extension of the Hopkinson's law:
\begin{equation}
    \Phi\mathcal{R}=nI_{f}+I_{\text{Hall},\varphi},
    \label{HopEq}
\end{equation}
where $n$ is the number of turns in the coil, $\Phi$ is the magnetic flux across the core section and $\mathcal{R}$ is the reluctance associated to the core (which basically depends on its cross section $A$ and average length $l$, i.e. $\mathcal{R}=l/\mu A$ \cite{zahn19791}). Here, we have identified $I_{Hall,{\varphi}}$ as the azimuthal current arising from the surface Hall current density (it is obtained from expression (\ref{SHC}) after integrating along the solenoid longitudinal direction). Since Eq. (\ref{HopEq}) relies on the constitutive relations (\ref{ECT1}) and (\ref{ECT2}), it holds for non-topological material as well, upon adhering to the condition that $\theta$ just changes in the interfaces. Let us emphasize that this result has not been reported before to the best of our knowledge. 

We can go further and consider the general scenario of an ideal transformer: two coils wound on a magnetic core such that all the magnetic flux of one coil links the other (for instance, the primitive transformer pictured by Fig. (\ref{Fig1Setup}.d)). Let us assume the coils (also called the primary and secondary wingdings) consist of $n_{1}$ and $n_{2}$ turns, and are fed by (free) currents $I_{f}^{(1)}$ and $I_{f}^{(2)}$ flowing in opposite direction. Hence, from the extended Hopkinson's law (\ref{HopEq}) follows 
\begin{equation}
\Phi\mathcal{R}
= 
n_{1}I_{f}^{(1)}- n_{2}I_{f}^{(2)}+I_{\text{Hall},\varphi}^{(1)}-I_{\text{Hall},\varphi}^{(2)},
    \label{HopEq2}
\end{equation}
where $I_{\text{Hall}\varphi}^{(i)}$ with $i=1,2$ are the surface Hall currents due to the magnetic fields generated by each coil (and $\mathcal{R}$ is the reluctance associated to the permeable core of arbitrary geometry). After replacing (\ref{HopEq2}) in the integral expression of the Faraday's law (i.e. $\epsilon^{(i)}=-\frac{d\Phi}{dt}$ with $i=1,2$), we obtain the induced electromotive forces across each coil, namely $\epsilon^{(i)}$, and thus, realize that the surface Hall currents give rise to an additional self and mutual inductions. Concretely, we show that the electromotive force of the primary winding in the magnetoquasitatic regime can be expressed as follows 
\begin{equation}
\epsilon^{(1)}=\big(L_{0}^{(1)}+L_{\text{ME}}^{(1)}\big)\frac{d I_{f}^{(1)}}{dt}- \big(M_{0}+M_{\text{ME}}\big)\frac{dI_{f}^{(2)}}{dt}, \label{FEMIPT}
\end{equation}
and similarly for the secondary winding after switching $1$ by $2$. Here, $L_{0}^{(i)}$ and $M_{0}$ are the self- and mutual- induction coefficients of the magnetic circuit in absence of the ME coupling. Essentially, by considering effects up to second-order in the ME polarizability, we have found out that: on one hand, the ME response can be cast in the form of self- and mutual- induction coefficients $L_{\text{ME}}^{(i)}$ and $M_{\text{ME}}$, respectively; and other hand, the impact of the radiative effects (which is characterized by the coefficient $N_{\text{rad}}$) is negligible at first-order in the magnetoquasistatic approximation despite of the ME effect. As a consequence, the primitive transformer endowed with the ME response (even combined with a strong bulk magnetization) retrieves an effective electromotive force that is identical to the conventional situation up to second order in $r_{2}\mu_{0}\omega_{0}\alpha_{0}\ll 1$ (see Eq.(\ref{EEMTFLSME1})). Let us emphasize that the derivation of Eq. (\ref{FEMIPT}) is based on an extensive perturbative analysis presented in Secs. \ref{SecLFR} and \ref{Sec_MCC}. 

\begin{figure}[t]
    \includegraphics[width=0.5\textwidth]{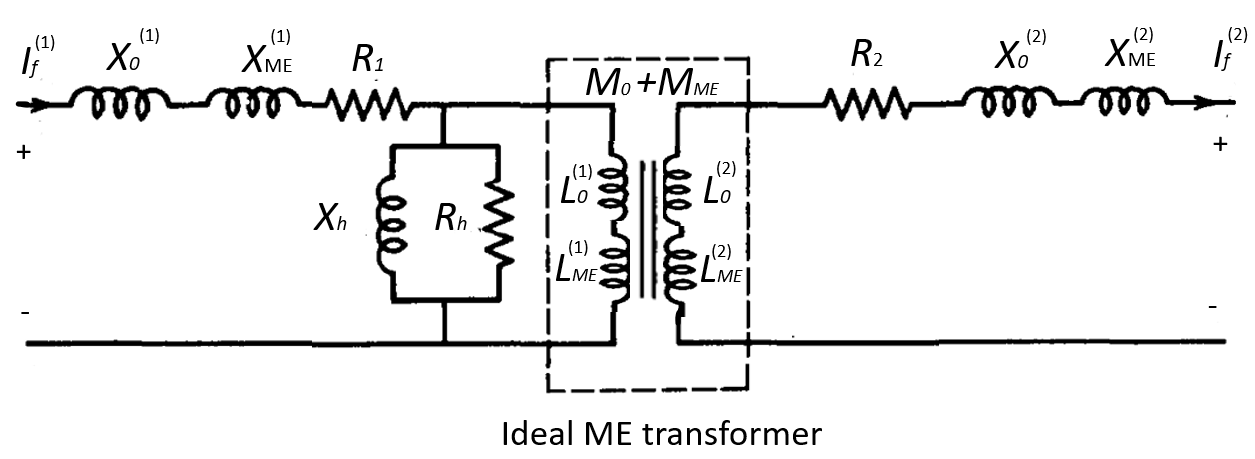}
  \caption{Diagram of a real ME transformer, composed of either a topological or an ordinary ME material, in the magnetoquasistatic domain. The influence of the ME response is explicitly shown in comparison with the conventional situation: $L_{ME}^{(i)}$, $M_{ME}$ and $X_{ME}^{(i)}$ denote the self-induction coefficients, mutual coefficient and leakage reactance relying on the ME response, respectively; while $L_{0}^{(i)}$, $M_{0}$ and $X_{0}^{(i)}$ correspond to the conventional situation in absence of the ME response. Notice that $R_{i}$ represent the electrical resistance associated to the conducting coils. \label{Fig1HS}} 
\end{figure}

To complete our treatment of a feasible transformer, we also estimate the magnetic leakage that represents all the magnetic flux spreads in the free space surrounding the coils. Concretely, in Sec \ref{Sec_RealT} we estimate the leakage reactances, namely $X^{(i)}$ with $i=1,2$, accounting for the magnetic flux produced by the primary winding that does not in the first and secondary coils (or viceversa); and find that it largely coincides with the conventional scenario: that is, $X^{(i)}\approx X_{0}^{(i)}+X_{\text{ME}}^{(i)}$ where $X_{\text{ME}}^{(i)}$ represents, at most, a perturbative correction of second order (in $r_{2}\mu_{0}\omega_{0}\chi \ll 1$) to the conventional magnetic leakage reactance $X_{0}^{(i)}$. That is, the magnetic flux remains practically confined within the permeable core for a weak ME polarizability as well. By collecting all these previous results, we finally come up to the transformer equivalent magnetic circuit composed of either a topological or an ordinary ME material, which is shown in Fig. (\ref{Fig1HS}). Compared with the conventional situation, it turns out that the main benefit of using TI insulators displaying the TME response resides on the fact that the core has virtually an insulating bulk preventing the emergence of eddy currents. That is, the topological magnetic circuit represents a more energetically efficient scheme for power converters as minimize the Joule heating arising in electrical manipulation \cite{tian20171}. Let us emphasize that diagram (\ref{Fig1HS}) is derived from a perturbative analysis that holds whenever the transformer operate within the quantum mechanical domain (which was detailed in the previous section) where $\theta$ is uniform in the bulk and quantized. Notice that in our analysis we have ignored the leakage effects due to the nonlinear magnetization of the core  \cite{liang20211} (e.g. the hysteresis losses), since we would need an additional microscopic model characteristic of the ME material to estimate its impact on the performance of the magnetic circuit, which is is beyond of the scope of the present work. Instead, in the magnetic circuit diagram we contemplate, as usual, a nonlinear inductive reactance $X_{h}$ \cite{zahn19791,wagemakers20171} and an electric resistance $R_{h}$ that only contains the power dissipated in traversing the hysteresis loop over a cycle (recall that we ignore the Joule heating owing the eddy currents would be suppressed), which could be experimentally assessed for certain ME materials \cite{fusil20141,cherifi20141,liang20211,tian20171}. 

Additionally, we have examined the opposite scenario for ME materials beyond the time-reversal-invariant TIs and AIs, we just require that $\theta$ only changes in the interfaces. In the domain of relatively high frequencies (i.e. $\frac{r_{2}\omega_{0}}{c_{i}}\gg  1$) and virtually large values of the ME susceptibility (e.g. $r_{2}\mu_{0}\omega_{0}\chi\gg  1$), we have discovered that the ME response diminishes significantly the electromagnetic fields, which in turn implies that the self-induction coefficient and the time-averaged electromagnetic energy get arbitrarily small. In contrast to the conventional situation, the generation of highly-intense electromagnetic fields becomes energetically expensive in the presence of significant ME couplings, as well as it restricts the operating frequency range to frequencies sufficiently small in comparison with the strength of the ME response. For instance, our treatment predicts that the operating frequencies of 2D multiferroic heterostructures could be substantially limited to frequencies below the very low frequency domain (recall that $\chi\sim 4.8\times 10^{3}\sqrt{\frac{\varepsilon_{0}}{\mu_{0}}}$ \cite{ying20221}). Let us mention that a similar drawback is manifested by several ME tunable inductors consisting of Metglass/PZT composites, instead an eddy current screening effect yields a monotonic decreasing of the inductance with frequency \cite{su20161}.  

Finally, we also touch upon the solenoid actuator in the magnetoquasistatic regime in Sec. \ref{Sec_FMHall}, where we provide an approximate estimation of the magnetic force strength that arises as a consequence of the surface Hall currents: this also represents a second-order correction in the ME susceptibility to the conventional situation (which is consistent with previous discussions).

\section{Long solenoid inductor}\label{Sec_BLS}

To provide a comprehensive analysis of the ME contribution on magnetic coupled circuits, it is convenient to start studying the electromagnetic fields for the scenarios of the bilayer long solenoid pictured by Fig. (\ref{Fig1Setup}). In this section, we perform an extensive numerical study of the self-induction coefficient, time-averaged electromagnetic energy and time-averaged power radiation for non-permeable media (i.e. $\mu_{1,r}=\mu_{2,r}=1$). We further provide several analytical results for both the low- and high- frequencies regimes, as well as for weak and strong ME responses. 

Although the returning fields outside the winds of the solenoid will generate the so-called fringe fields, these will be significantly small in comparison with the field inside the solenoid as long as $l \gg r_{2}$. Here, we work within the domain where the fringe fields can be neglected \cite{harmon19911,thevenet19991,yang20211,lin20211}, such that the solenoid can be treated as infinitely long. The symmetry of the system suggests that we should employ the cylindrical coordinates $(r,\varphi, z)$ were the $z$-axis coincides with the solenoid's central axis, whereas its center is taken as the origin. Hence, an alternating, uniform transverse surface current at the interface $r=r_2$ is considered (notice that we take a null external charge distribution, i.e. $\rho_{f}(\vect r,t)=0$), i.e. 
\begin{equation}
    \vect J_{f}(\vect r,t)=\frac{n I_{f}(t)}{l}\delta(r-r_2)\ \hat \varphi,
    \label{EQJ}
\end{equation}
which represents a coil composed of $n$ turns and supplied with the alternating current previously introduced. As anticipated, it is important to note that, the electrical permittivity, magnetic permeability and ME polarizability $\theta(\vect r)$ are piecewise constant functions along the radial distance, e.g.
\begin{equation}
   \theta(\vect r) =\theta_{1}+(\theta_2 - \theta_1)\Theta( r- r_1)+(\theta_3 - \theta_2)\Theta( r- r_{2}).
\label{AAG}
\end{equation}
with $\Theta(r)$ being the step-Heaviside function. To the best of our knowledge, there is no previous works that treat the ac bilayer solenoid described in the standard Maxwell electrodynamics (all known result restricts to the single-layer scenario \cite{harmon19911,zangwill20121,templin19951}), so the present paper proves beneficial in this aspect as well.  Owing to the current density (\ref{EQJ}) follows a harmonic function, we may expect the electromagnetic fields manifest a harmonic dependence with in time characterized by $\omega_{0}$ as well. Hence, it is convenient to introduce the $\omega$-variable Fourier transform $\hat{f}$ of a time-dependent function, e.g.
\begin{equation*}
  f(\vect r,t)=\frac{1}{2\pi} \int_{-\infty}^{+\infty}d\omega \  \hat{f}(\vect r,\omega) e^{-i\omega t},
 \label{FouTG}
\end{equation*}
and its complex conjugate $\hat{f}(\vect r,\omega)=\hat{f}^{\dagger}(\vect r,\omega) $. 

To tackle the electromagnetism problem we switch to the electromagnetic potentials $\phi(\vect r, t)$ and $\vect A(\vect r,t)$, where $\vect B=\nabla\times A$ and $\vect E=-\nabla\phi-\frac{1}{c}\partial_{t}\vect A$. Since we are dealing with time-dependent electromagnetic fields, we solve the extended Maxwell equations in the Lorenz gauge \cite{jackson20121}. The cylindrically symmetric geometry implies that both $\vect E$ and $\vect B$ have no radial components and that their magnitudes are independent of the angular degree of freedom (see the scenario with conventional insulators \cite{rivera20021,harmon19911,thevenet19991}). Collecting these
observations we may rewrite the electromagnetic potentials as follows,
\begin{eqnarray}
\phi(\vect r,t)&=&\phi(t), \label{PAsntph} \\
\vect A(\vect r,t)&=&A_{\varphi}(r,t)\hat \varphi+A_{z}(r,t) \hat z. \label{PAsntA}
\end{eqnarray}
We note that the $z$-axis (or axial) component of the vector potential cancels in the conventional solenoid excited by an alternating current \cite{abbott19851,zangwill20121}, which means that $A_{z}$ arises exclusively from the ME response. More precisely, we anticipate that the latter arises from the bulk magnetization due to the $\theta$-term in (\ref{ECT1}), which shall be refereed to as $\theta$-magnetization.

By plugging the ansatz (\ref{PAsntph}) and (\ref{PAsntA}) into the wave equation for the electromagnetic potentials obtained from Eqs. (\ref{ME1})-(\ref{ME4}) after setting Eq. (\ref{EQJ}), we get the simple equation of motion for the scalar potential, 
\begin{equation*}
\frac{\partial^2 \phi}{\partial t^2}=0   
\end{equation*}
which yields the trivial solution, i.e. $\phi(t)=\phi_{1}t+\phi_0$. Combined with the fact that the radial contribution to the vector potential wave equation, one finds that $\phi_{1}=0$, and thus $\phi(t)=\phi_0$. Without loss of generality, we shall assume $\phi_0=0$ so the solenoid electromagnetic fields are fully determined by the vector potential, as similarly occurs in the conventional Maxwell electromagnetism in absent of charge distributions \cite{jackson20121}. The vector potential equations of motion can be readily expressed in terms of its Fourier transform for  $r\neq r_{1}, r_{2}$,
\begin{widetext}

\begin{eqnarray}
\bigg(\frac{\partial^2}{\partial r^2}+\frac{1}{r}\frac{\partial}{\partial r} -k^2-\frac{1}{r^2} \bigg)\hat A_{\varphi}+\frac{\alpha_{0}\mu_{r}\mu_{0}}\omega{\pi}\frac{\partial \theta}{\partial r}\hat A_{z}+\frac{\mu_{r}}{r}\frac{\partial \mu_{r}^{-1}}{\partial r}\frac{\partial(r \hat A_{\varphi})}{\partial r}&=&0, \label{EoMAP} \\
\bigg(\frac{\partial^2}{\partial r^2}+\frac{1}{r}\frac{\partial}{\partial r}-k^2-\frac{1}{r^2}\bigg)\hat  A_{z}-\frac{\alpha_{0}\mu_{r}\mu_{0}\omega}{\pi}\frac{\partial \theta}{\partial r}\hat A_{\varphi}+\mu_{r}\frac{\partial \mu_{r}^{-1}}{\partial r}\frac{\partial \hat A_{z}}{\partial r}&=&0, \label{EoMAZ}
\end{eqnarray}
where we have identified the wave number of the electromagnetic waves, i.e. $k=\omega\sqrt{\varepsilon_{r}\mu_{r}}/c$. Furthermore, by starting from Eqs. (\ref{EBoundCond1})-(\ref{EBoundCond4}) we obtain the boundary conditions for the vector potential in the cylindrical coordinate system, that is
\begin{align}
     \hat A_{\varphi}(r_{i}^{-},\omega)=& \hat A_{\varphi}(r_{i}^{+},\omega), \label{ABoundCond1} \\
    \hat  A_{z}(r_{i}^{-},\omega)=& \hat A_{z}(r_{i}^{+},\omega), \label{ABoundCond2}
\end{align}
and for $\omega=\omega_{0}$ (or the complex conjugate for $\omega=-\omega_{0}$),
\begin{align}
    \frac{1}{\mu_{i}}\bigg(\frac{1}{r}\frac{\partial (r \hat A_{\varphi})}{\partial r}\bigg)\bigg|_{r=r_{i}^{-}}- \frac{1}{\mu_{i+1}}\bigg(\frac{1}{r}\frac{\partial(r\hat  A_{\varphi})}{\partial r}\bigg)\bigg|_{r=r_{i}^{+}}&=-\frac{ nI_{0}}{l} \delta_{i2}+i\frac{\omega_{0}}\alpha_{0}{\pi}(\theta_{i+1}-\theta_i)\hat A_{z}(r_{i},\omega), \label{ABoundCond3} \\
    \frac{1}{\mu_{i}}\bigg(\frac{\partial \hat A_{z}}{\partial r}\bigg)\bigg|_{r=r_{i}^{-}}-\frac{1}{\mu_{i+1}}\bigg(\frac{\partial \hat A_{z}}{\partial r}\bigg)\bigg|_{r=r_{i}^{+}}&=-i\frac{\omega_{0}}\alpha_{0}{\pi}(\theta_{i+1}-\theta_i)\hat A_{\varphi}(r_{i},\omega), \label{ABoundCond4} 
\end{align}
where $\mu_{a}=\mu_{a,r}\mu_{0}$ is the permeability of the medium and $\delta_{ij}$ represents the Kronecker delta. By paying attention to Eqs. (\ref{EoMAP})-(\ref{ABoundCond4}), one may realize that the ME effect disappears in the strict magnetostatic regime (i.e. $\omega\rightarrow 0$). This point will be further discussed below. 

As anticipated in Sec. \ref{SecTME}, the contribution of a piecewise constant ME coupling in the homogeneous modified Maxwell's equations turns out to be null despite the constitutive relations (\ref{ECT1}) and (\ref{ECT2}) explicitly depend of $\theta$. Here, it is important to realize that the TME effect just give rises to a surface contribution \cite{armitage20191}, that is the surface Hall current (\ref{SHC}) (as the gradient of the ME coupling (\ref{AAG}) turns into a combination of Dirac delta functions). Recall that the $\theta$-electrodynamics  (\ref{ME1})-(\ref{ME4}) in time-reversal-invariant TIs and AIs manifest at surfaces and interfaces \cite{nogueira20221,sekine20211}. 
Beside, the electrical permittivity and magnetic permeability are piecewise constant functions as well, i.e.
\begin{equation}
   \varepsilon(r,\omega)
 = \left\{
 \begin{array}{cl}
    \varepsilon_0\varepsilon_{1,r}(\omega),  & r\leq r_1, \\
    \varepsilon_0\varepsilon_{2,r}(\omega),  & r_1 < r<r_{2}, \\
    \varepsilon_0,  & r_2 \le r. 
 \end{array}
 \right. \\\\\
    \mu(r)
 = \left\{
 \begin{array}{cl}
    \mu_0\mu_{1,r}(\omega),  & r\leq r_1, \\
    \mu_0\mu_{2,r}(\omega),  & r_1 < r<r_{2}, \\
    \mu_0,  & r_2 \le r. 
 \end{array}
 \right. \\\\\
  k(r,\omega)
 = \left\{
 \begin{array}{cl}
    k_1=\frac{\omega\sqrt{\mu_{1,r}\varepsilon_{1,r}(\omega)}}{c}=\frac{\omega}{c_{1}(\omega)},  & r\leq r_1, \\
    k_2=\frac{\omega\sqrt{\mu_{2,r}\varepsilon_{2,r}(\omega)}}{c}=\frac{\omega}{c_{2}({\omega})},  & r_1 < r<r_{2}, \\
    k_3=\frac{\omega}{c}=\frac{\omega}{c_{3}},  & r_2 \le r,
 \end{array}
 \right.
 \label{EQPPERWN}
\end{equation}
where $\varepsilon_{i,r}(\omega)$ is given by Eq. (\ref{EQDISRLF}) for the ME media, as stated in Sec. \ref{SecTME}. Notice that we must distinguish between the inner and outer regions of the bilayer solenoid: that is, $\varepsilon_{1,r}(\omega)=\varepsilon_{r}(\omega)$ and  $\varepsilon_{2,r}(\omega)=\tilde\varepsilon_{r}$ for the first arrangement,  $\varepsilon_{2,r}(\omega)=\varepsilon_{r}(\omega)$ and  $\varepsilon_{1,r}(\omega)=\tilde\varepsilon_{r}$ for the second arrangement, and $\varepsilon_{1,r}(\omega)=\varepsilon_{2,r}(\omega)=\varepsilon_{r}(\omega)$ for the third arrangement of the TI. To solve Eqs. (\ref{EoMAP})-(\ref{EoMAZ}) (under the boundary conditions (\ref{ABoundCond1})-(\ref{ABoundCond4})) 
we then take advantage of these features: these imply that the second and third terms on the right hand side of Eqs. (\ref{EoMAP})-(\ref{EoMAZ}) cancel for $r\neq r_{1},r_{2}$; yielding the usual Bessel equation found in the conventional solenoid \cite{harmon19911,thevenet19991}. As it is well known \cite{zangwill20121,mcdonald19971,templin19951,harmon19911,thevenet19991}, the solution of the latter is a combination of both the Bessel and Hankel functions of the first kind, $J_{1}(x)$ and $H_{1}^{(1)}(x)$. Based on this observation, we propose the solutions for $\omega=\omega_{0}$ (or the complex conjugate for $\omega=-\omega_{0}$)
\begin{equation}
   \hat A_{\varphi}(r,\omega_{0})
 = \left\{
 \begin{array}{cl}
    a_{\varphi}(r_{1},r_{2},\omega_{0})J_{1}(k_{1} r),  & r\leq r_1, \\
    c_{\varphi}(r_{1},r_{2},\omega_{0})J_{1}(k_{2} r)+c_{\varphi}(r_{1},r_{2},\omega_{0})H_{1}^{(1)}(k_{2} r),  & r_1 < r<r_{2}, \\
   d_{\varphi}(r_{1},r_{2},\omega_{0})H_{1}^{(1)}(k_{3} r),  & r_2 \le r. 
 \end{array}
 \right. \label{ESAPhi}
\end{equation}
and 
\begin{equation}
     \hat A_{z}(r,\omega_{0})
 = \left\{
 \begin{array}{cl}
    a_{z}(r_{1},r_{2},\omega_{0})J_{1}(k_{1} r),  & r\leq r_1, \\
    c_{z}(r_{1},r_{2},\omega_{0})J_{1}(k_{2} r)+c_{z}(r_{1},r_{2},\omega_{0})H_{1}^{(1)}(k_{2} r),  & r_1 < r<r_{2}, \\
   d_{z}(r_{1},r_{2},\omega_{0})H_{1}^{(1)}(k_{3} r),  & r_2 \le r. 
 \end{array}
 \right.  \label{ESAz}
\end{equation}
where we have imposed that both $\hat A_{z}(r,\omega_{0})$ and $\hat A_{\varphi}(r,\omega_{0})$ must converge for $r<r_{1}$ and $r>r_{2}$, and there is none initial wave propagating inward toward the origin from  the infinity \cite{templin19951,harmon19911,zangwill20121,mcdonald19971} (physically, the radiation field will consist of outgoing waves). Here, we have also introduced the coefficients $a_{\varphi/z}$, $b_{\varphi/z}$ and $c_{\varphi/z}$; which are fully determined by patching the above solutions together via the boundary conditions (\ref{ABoundCond1})-(\ref{ABoundCond4}) \cite{chew19951}. This is extensively illustrated in App. \ref{AppBLSSol}, see Eqs. (\ref{LASE}) and (\ref{DLQUMI}), where it is shown that these coefficients also depend of Bessel and Hankel functions of the first kind $J_{0}(x)$, $H_{0}^{(1)}(x)$, $J_{2}(x)$, and $H_{2}^{(1)}(x)$ evaluated at the solenoid interfaces.  Let us recall that we concentrate on the frequency domain away from the relaxation peak located at the resonance frequency $\omega_{R}$ such that strong absorption effects are negligible and $\epsilon_{r}(\omega)$ is real and positive (otherwise our preceding treatment is no longer valid): this domain is $\omega_{0}<\omega_{R}$ and $\omega_{0}>\sqrt{\omega_{R}^2+\omega_{e}^2}$.

\end{widetext}

\begin{figure*}[ht]
\centering
\includegraphics[scale=0.52]{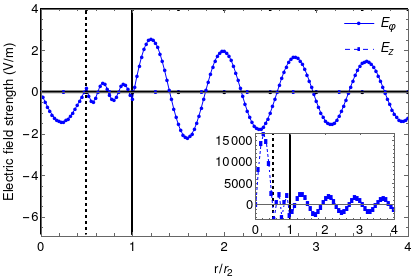}
\includegraphics[scale=0.54]{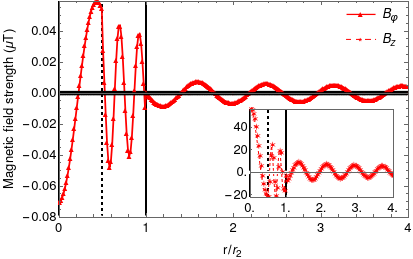}
\includegraphics[scale=0.52]{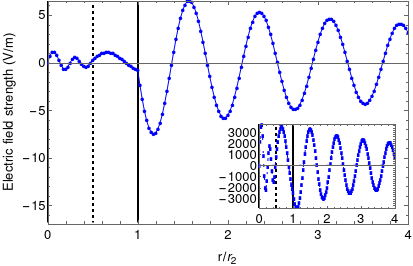}
\includegraphics[scale=0.54]{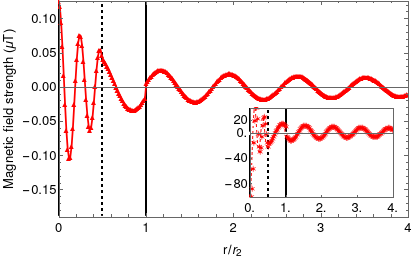}
\includegraphics[scale=0.52]{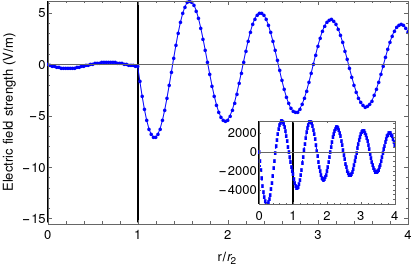}
\includegraphics[scale=0.54]{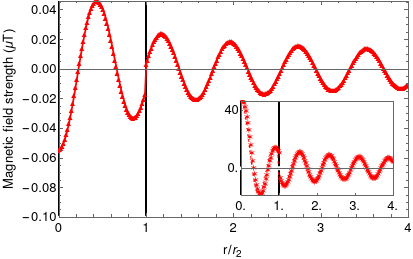}
\caption{(color online) Electromagnetic fields as a function of the radial length for the three arrangements of bilayer solenoid inductors sketched in Fig. (\ref{Fig1Setup}): TI-NI (upper), NI-TI (central), TI-TI (down). The blue solid and dashed lines represent respectively the electric fields $E_{\varphi}$ and $E_{z}$, while the red solid and dashed lines represent respectively the magnetic fields $B_{\varphi}$ and $B_{z}$. In all the pictures, the vertical dashed and solid lines illustrate the inner regions with $r_{1}$ and $r_{2}$, respectively. The parameters were chosen as $\omega_{0}=24.1$ THz, $t=\pi/8\omega_{0}$ and $nI_{0}/l=1$ A/m, whereas the solenoid inner and outer radius were taken $r_{1}=0.05$ mm and $r_{2}=0.1$ mm, respectively.
  \label{Fig3}} 
\end{figure*}

Once we have done the inverse Fourier transform of (\ref{ESAPhi})-(\ref{ESAz}) these can be replaced in (\ref{PAsntA}), which yields the following expression of the vector potential in the ac bilayer solenoid (notice that the electromagnetic fields are considered to be zero in absence of the exciting current $I_{f}$),
\begin{align}
\vect A(r,t)&=\Big[\text{Re} \ \hat A_{\varphi}(r,\omega_{0})\cos(\omega_{0} t)
\nonumber \\
& \quad \quad  +\text{Im}\ \hat A_{\varphi}(r,\omega_{0}) \sin(\omega_{0} t)\Big]\hat{\varphi} \nonumber \\
& \, +\Big[\text{Re} \ \hat A_{z}(r,\omega_{0})\cos(\omega_{0} t)\nonumber \\
&\quad \quad  +\text{Im}\ \hat A_{z}(r,\omega_{0}) \sin(\omega_{0} t)\Big]\hat z, 
\label{Amu01}
\end{align}
where we have made used of the Schwarz reflection principle valid for axionic materials \cite{crosse20151}, i.e. $\hat A(r,\omega_{0})=\hat A^{\dagger}(r,-\omega_{0})$ for all $r$. We next use the expressions of the electric and magnetic fields in term of the vector potential and substitute (\ref{Amu01}), after some manipulation we get
\begin{align}
\vect E(r,t) 
=
\omega_{0} 
& \Big(
 \Big[\text{Re} \ \hat A_{\varphi}(r\omega_{0})\sin(\omega_{0} t) \nonumber \\
& \quad \quad
-\text{Im}\ \hat A_{\varphi}(r,\omega_{0}) \cos(\omega_{0} t)\Big]\hat{\varphi} \nonumber \\
& 
+\Big[\text{Re} \ \hat A_{z}(r,\omega_{0})\sin(\omega_{0} t)\nonumber \\
& \quad \quad
-\text{Im}\ \hat A_{z}(r,\omega_{0}) \cos(\omega_{0} t)\Big]\hat z\Big), \label{ExEf}
\end{align}
and 
\begin{align}
\vect B(r,t) 
&=-\frac{\partial }{\partial r}\Big[\text{Re} \ \hat A_{z}(r,\omega_{0})\cos(\omega_{0} t) \nonumber \\
& \qquad \qquad 
+\text{Im}\ \hat A_{z}(r,\omega_{0}) \sin(\omega_{0} t)\Big] \hat{\varphi} \nonumber \\ 
&\quad 
+\Big[\frac{1}{r}\Big(\text{Re} \ \hat A_{\varphi}(r,\omega_{0})\cos(\omega_{0} t)\nonumber \\
&\qquad \qquad 
+\text{Im}\ \hat A_{\varphi}(r,\omega_{0}) \sin(\omega_{0} t) \Big)  \nonumber \\
&\qquad \qquad 
+\frac{\partial}{\partial r}\Big(\text{Re} \ \hat A_{\varphi}(r,\omega_{0})\cos(\omega_{0} t)\nonumber\\
&\qquad \qquad 
+\text{Im}\ \hat A_{\varphi}(r,\omega_{0}) \sin(\omega_{0} t) \Big)\Big]\hat z.  \label{BxBf}
\end{align}
We shall show in Sec. \ref{SecLFR} that these expressions return the electric and magnetic fields of the vacuum solenoid in the magnetostatic limit when $\varepsilon_{1}=\varepsilon_{2}=\varepsilon_0$, as expected. Furthermore, even though it is not shown here, from Eqs. (\ref{Amu01})-(\ref{BxBf}) one also recovers the conventional solution of the vacuum solenoid excited with the alternating current density (\ref{EQJ}) when neglecting the ME effect \cite{zangwill20121,mcdonald19971,templin19951,harmon19911}, i.e. $\theta_{i}\rightarrow 0$ for $i=1,2$. 

From Eqs. (\ref{ExEf}) and (\ref{BxBf}) follows that the ME response gives rise to both an additional magnetization and polarization that corresponds to an azimuthal magnetic field $B_{\varphi}$ as well as a longitudinal electric field $E_{z}$ (beside the axial magnetic $B_{z}$ and circumferential electric $E_{\varphi}$ fields). The origin of these new components of the electromagnetic fields can be traced back to the non-dissipative surface Hall currents that arise at the cylindrical surfaces of the TI layers, indicating that they are a direct consequence of a dynamical surface response. More specifically, it turns out that longitudinal surface Hall currents $I_{\text{Hall}, i}^{z}$ (with $i=1,2$) are responsible for the generation of $B_{\varphi}$: this will be explicitly shown for low frequencies and weak ME responses (that is $r_{2}\omega_{0}/c_{3}\gg1$ and $r_{2}\mu_{0}\omega_{0}\alpha_{0}\ll 1$) in Sec. \ref{SecLFR}, where we provide reduced expressions of the azimuthal magnetic field (see Eqs. (\ref{BpMQA1}) and (\ref{BpMQA2})). These currents are obtained from the response current density (\ref{SHC}) after replacing the electric field (\ref{ExEf}), which yields
\begin{align}
     I_{\text{Hall},1}^{z}(t)&=2\pi r_{1}\alpha_{0}(\theta_{2}-\theta_{1})E_{\varphi}(r_{1},t), \label{DCHSZ1} \\
    I_{\text{Hall},2}^{z}(t)&=2\pi r_{2}\alpha_{0}(\theta_{3}-\theta_{2})E_{\varphi}(r_{2},t), \label{DCHSZ2}
\end{align}
where $I_{\text{Hall},1}^{z}(t)$ and $I_{\text{Hall},2}^{z}(t)$ flow along the solenoid axis at the inner and outer interfaces with radius $r_{1}$ and $r_{2}$, respectively. Observe that the longitudinal surface Hall current is set up by the azimuthal electric field, which is self-consistently induced by $B_{z}$ according to the induction Faraday’s law (\ref{ME3}). Still, we can go further and identify $B_{\varphi}$ as the responsible of the axial electric field by appealing to Faraday’s law as well. That is, the changing magnetic flux created by $B_{\varphi}$ (recall that $\partial E_{z}/\partial r=\partial B_{\varphi}/\partial t$ follows from Eq.(\ref{ME3})), will induce $E_{z}$ as a consequence of longitudinal surface Hall currents. 

Additionally, azimuthal surface Hall currents $I_{\text{Hall},i}^{\varphi}$
arise owing to the ME response. Following a similar procedure as before, we obtain
\begin{align}
    I_{\text{Hall},1}^{\varphi}(t)&=-\frac{\alpha_{0} l}{\pi }(\theta_{2}-\theta_{1})E_{z}(r_{1},t), \label{DCHSIP1} \\
    I_{\text{Hall},2}^{\varphi}(t)&=-\frac{\alpha_{0} l}{\pi}(\theta_{3}-\theta_{2})E_{z}(r_{2},t) \label{DCHSIP2},
\end{align}
where $I_{\text{Hall},1}^{\varphi}(t)$ and $I_{\text{Hall},2}^{\varphi}(t)$ flow around the solenoid circumference at the inner $r_{1}$ and outer $r_{2}$ radius, respectively. Unlike $I_{\text{Hall},i}^{z}(t)$ ($i=1,2$), the azimuthal surface Hall currents exclusively depend of the axial electric field.  As $I_{\text{Hall},i}^{z}$ give rises to $B_{\varphi}$, one may expect that $B_{z}$ can be regarded as the superposition of the axial magnetic fields produced by both the free $I_{f}$ and the azimuthal surface Hall current $I_{\text{Hall}}^{\varphi}$. Indeed, we show for small frequencies and weak ME responses in Sec.\ref{SecLFR} that $B_{z}$ takes the well-know expression of the quasistatic magnetic field of a conventional solenoid supplied by currents $I_{f}$ and  $I_{\text{Hall}}^{\varphi}$ (see Eqs. (\ref{BzMQAa}) and (\ref{BzMQAb})).  Combining together these previous results, we arrive to the following diagram that summarizes the interplay between the electromagnetic fields:
\begin{center}
$\begin{CD}
B_{z} @>\textbf{Electromagnetic induction}>(1)> E_{\varphi} \\
@A\text{$\theta$-magnetization}A(4)A @V(2)V\text{$\theta$-magnetization}V\\
E_{z} @<(3)<\textbf{Electromagnetic induction}< B_{\varphi}
\end{CD}$ 
\end{center}
The above diagram can be intuitively understood from the ME response (\ref{ECT1}): from the latter directly follows the induced magnetization at the steps (2) and (4) (where surface Hall currents identify with magnetization currents), while the steps (1) and (3) are explained by the Faraday's law (at least, in the magnetoquasistatic regime). From this diagram is clear the dynamical mechanism behind the ME effect.

Before proceeding to the numerical evaluation of Eqs. (\ref{ExEf}) and (\ref{BxBf}), it is interesting to realize that the ME effect is expected to manifest more intensively in the NI-TI setup than in the TI-NI and TI-TI configurations. Going back to the setup sketched in Fig. (\ref{Fig1Setup}), one may see that the TI-NI and TI-TI configurations will display the longitudinal $I_{\text{Hall},i}^{z}(t)$ and azimuthal $I_{\text{Hall},i}^{\varphi}(t)$ currents flowing in the inner or outer cylindrical surface (with radius $r=r_{i}$ for $i=1,2$), respectively (since the cylindrical TIs has a single curved surface); whereas, the NI-TI setup will exhibit $I_{\text{Hall},i}^{z}(t)$ and $I_{\text{Hall},i}^{\varphi}(t)$ running in both the inner and  outer curved surfaces (i.e. $i=1,2$).

\begin{figure*}
\centering
\includegraphics[scale=0.32]{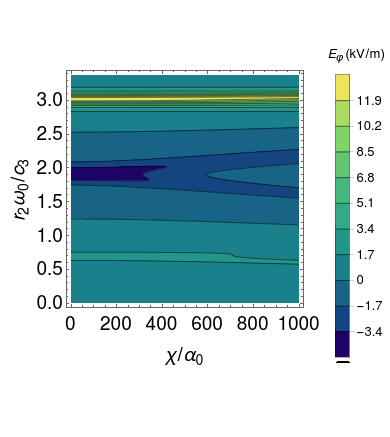}
\includegraphics[scale=0.32]{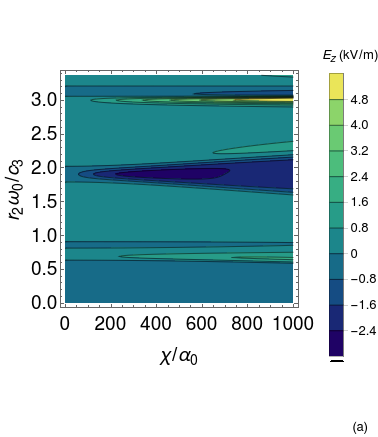}
\includegraphics[scale=0.32]{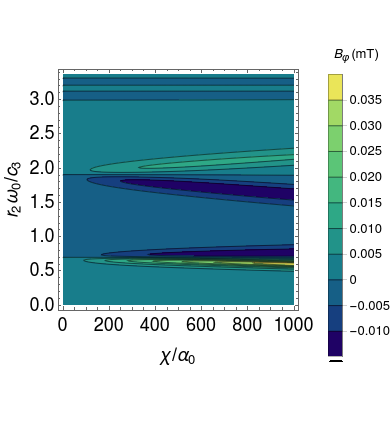}
\includegraphics[scale=0.32]{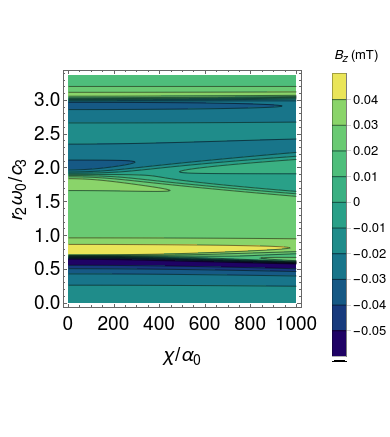}
\includegraphics[scale=0.32]{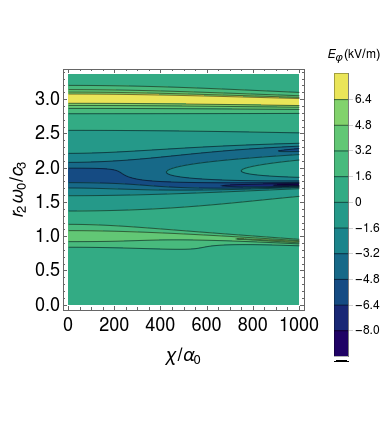}
\includegraphics[scale=0.32]{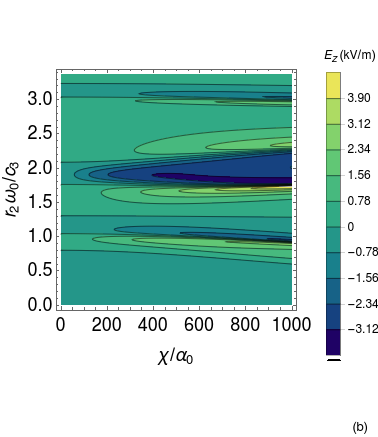}
\includegraphics[scale=0.32]{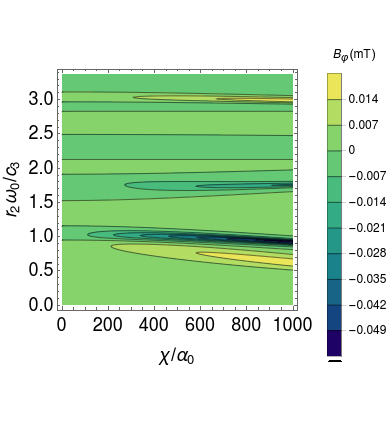}
\includegraphics[scale=0.32]{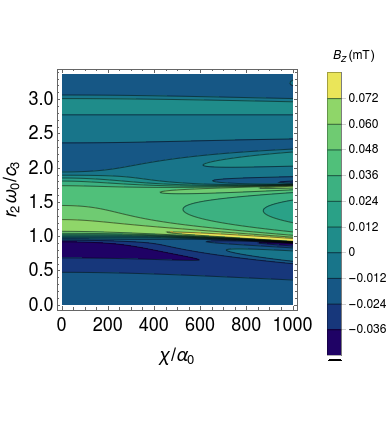}
\includegraphics[scale=0.32]{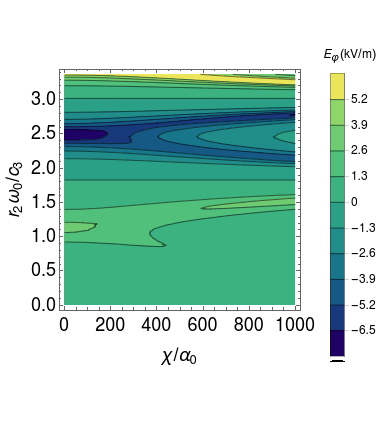}
\includegraphics[scale=0.32]{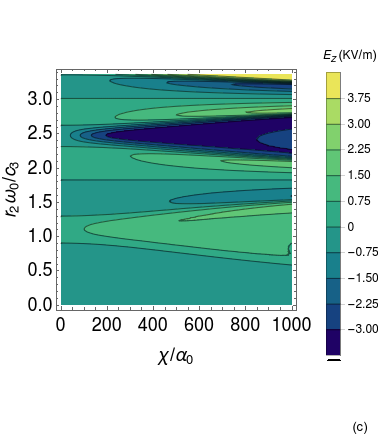}
\includegraphics[scale=0.32]{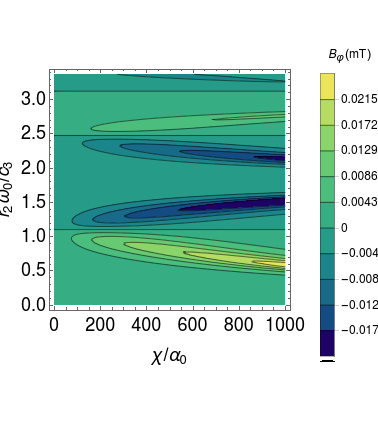}
\includegraphics[scale=0.32]{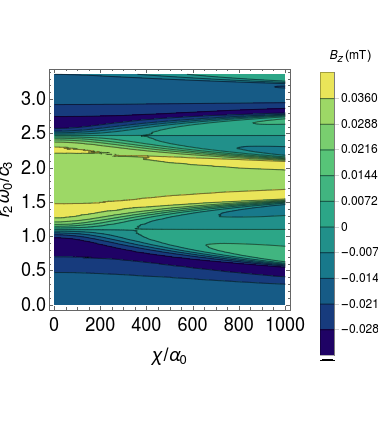}
\caption{(color online). Contour plots of electromagnetic fields as a function of the exciting frequency and the ME susceptibility for the three arrangements of bilayer solenoid inductor sketched in Fig. 
 (\ref{Fig1Setup}): TI-NI (a), NI-TI (b), TI-TI (c). In all pictures, we consider $nI_{0}/l=1$ A/m, and the radial distance and time were fixed to $r=0.25r_{2}$, $t=\pi/\omega_{0}$, whereas the solenoid inner and outer radius were taken $r_{1}=0.5$ mm and $r_{2}=1$ mm, respectively. Recall that we have considered non-permeable media  (i.e. $\mu_{1,r}=\mu_{2,r}=1$). \label{Fig4}} 
\end{figure*}

\subsection{Numerical computation of electromagnetic fields}\label{Sec_NCEMF}

In order to compute electromagnetic fields for practical purpose scenarios, we will take the dielectric constant for the trivial insulator as $\tilde\varepsilon_{r}=12$, which have been extensively used in a number of works dealing with TIs in the THz frequency domain \cite{crosse20151,shoukat20161,sekine20211,crosse20171,franca20211,crosse20161,martin20191,qi20081,martin20181,nenno20201}, as well as we assume $\mu_{1,r}=\mu_{2,r}=1$ since a broad class of TI are non-magnetic \cite{medel20231,martin20181}. We also employ the international system of units (for which $\varepsilon_{0}=8,854\ 10^{-12}$ F/m and $\mu_{0}=4\pi \  10^{-7}$N/A$^2$) from now on. 

Let us first study the situation of three solenoid inductors composed of certain TI excited by a frequency $\omega_{0}=24.1$ THz (which is well embedded in the surface band gap of prototypical TIs). Shown in Fig. (\ref{Fig3}) are the electromagnetic fields for three arrangement of the TI as a function of the radial distance, denoted by $r$, in the millimeter length scale. Concretely, the longitudinal electric $E_{z}$ and azimuthal magnetic $B_{\varphi}$ components are depicted in the main frame, whereas the azimuthal electric $E_{\varphi}$ and longitudinal magnetic $B_{z}$ components are illustrated in the corresponding insets. Importantly, the amplitude of the former electromagnetic fields are three or four order of magnitude lower than the latter. In other words, the electromagnetic fields exclusively arising from the TME response remain perturbative, this point will be further discuss below. Although the magnitudes of the electromagnetic fields are barely modified, it can be also seen that their behaviour inside the solenoid substantially change from one arrangement to other of the TI layers. In particular, the azimuthal component of the magnetic field manifests abrupt variations at the TI surfaces (for instance, see at the inner radio $r_{1}=0.5r_{2}$ in the configuration NI-TI), which is a signature of $I_{\text{Hall}}^{\varphi}$ and $I_{\text{Hall}}^{z}$ since they are sources of the magnetic field, as previously pointed out.

Upon further inspection, we appreciate that $E_{z}$ and $B_{\varphi}$ are shifted in time by $\pi$ radians outside the solenoid in all studied setups, unlike $E_{\varphi}$ and $B_{z}$ which are in the same phase as in the conventional situation \cite{harmon19911,thevenet19991}. According to the expression of the Poynting vector (\ref{ESEM}), this is consistent with an outgoing radiation field, so that the ME effect contribute to an outward electromagnetic energy flow through the solenoid surface (with radius $r=r_{2}$). In Secs. \ref{SecLFR} and \ref{SecHFR} we show that the time-averaged of the power radiated by the solenoid depends on the bilayer arrangement. Inside the solenoid the situation slightly changes: $E_{z}$ and $B_{\varphi}$ as well as $E_{\varphi}$ and $B_{z}$ are $\pi/2$ radians out of phase in the time domain, which is consistent with the diagram about the formation of the electromagnetic fields discussed in the previous section. Moreover, this coincides with the standard vacuum solenoid  \cite{harmon19911,zangwill20121}.

In App. \ref{AppBLSSol} we also provide electromagnetic fields as functions of the radial distance and the ME polarizability, see Fig. (\ref{FigErMEC}). Essentially, we find that the electromagnetic components present a rich radial distribution in terms of $\chi$ in the NI-TI and TI-TI configurations.  For instance, by observing the amplitude of the axial $E_z$ and circumferential $B_\varphi$ components, it can be seen that they exhibit well-defined peaks inside the solenoid for middle values of the ME coupling coefficient. Additionally, $E_\varphi$ and $B_z$ exhibit either a decreasing or increasing behavior for varying $\chi$, which contrasts with the results retrieved by the TI-NI arrangement (as such components remain widely unmodified). This feature is particularly stressed for the axial component $B_{z}$, which suggest that the ME response is specially promoted by the NI-TI and TI-TI configurations. We also find that the amplitude of all the electromagnetic components eventually decay with the radial distance (that can be clearly observed in Fig. (\ref{Fig3})) independent of the ME coupling coefficient. Furthermore, it can be observed that the attenuation of the electromagnetic fields outside the solenoid is barely modified by increasing $\chi$. In particular, $B_z$ mainly accumulates inside the solenoid for all the studied bilayer configurations, which resemblances the concentrating ability of the vacuum solenoid \cite{yang20211}.

One may further appreciate from Fig. (\ref{FigErMEC}) that the circumferential $E_{\varphi}$ and axial $B_{z}$ components are three order of magnitude larger than $E_{z}$ and $B_{\varphi}$ when $\chi/\alpha_{0}$ is close to the unit (recall that $\chi= \alpha_{0}$ in time-reversal TIs and AIs  \cite{sekine20211}). As previously anticipated, this means that the ME effect upon electromagnetic fields are comparatively weak for typical values of the ME susceptibility manifested by TI. To deep into this question, we have also examined how the electromagnetic fields behave in the exciting frequency and ME polarizability parameter space. Concretely, in Fig. (\ref{Fig4}) we provide contour plots of electromagnetic fields as functions of $\omega_{0}$ and $\chi$ expressed in units of the fine-structure constant. Notice that the upper right corner of these plots (which represents THz frequencies and ME susceptibilities close to the fine-structure constant) corresponds to the results retrieved by TIs manifesting the TME effect, while the lower left corner (which regards the RF domain and the largest value of the ME susceptibility) reflects the results due to the heterostructure multiferroics mentioned at the beginning of Sec. \ref{SecTME}. A quick glance reveals that the electric and magnetic fields in all the situations are barely modified by the ME susceptibility for a current frequency close to zero (that is, $r_{2}\omega_{0}/c_{3}\ll1$), which stress out that the ME interaction play a role in the solenoid only when it is excited by alternating currents (recall that the electromagnetic phenomena studied here arises from a dynamical ME effect). One may also appreciate that electromagnetic fields exhibit an intricate behavior depending on the bilayer configuration: while all the electromagnetic fields in the TI-NI arrangement are largely uniform for varying $\chi$ up to certain frequency below $r_{2}\omega_{0}/c_{3}< 1.5$, the ME effect manifest at comparatively smaller frequencies in the NI-TI and TI-TI setups. Let us outline that the displacement electric effect becomes more intensive as we goes up vertically, whereas the ME response gets stronger as we move horizontally from the left to the right.

Fixing the frequency at middle values $r_{2}\omega_{0}/c_{3}\sim 1$ (which corresponds to practical frequencies around $0.3$ THz), it can be observed that most of electromagnetic components first rise for an initially increasing  ME coupling strength in the three configurations. For instance, $B_{z}$ initially takes on larger values with respect to the standard vacuum solenoid for a comparatively large $\chi$ (i.e. $\chi\sim 500\alpha_{0}$). The later can be intuitively understood by paying attention to $E_{z}$ and recalling that the azimuthal surface Hall currents $I_{\text{Hall},i}^{\varphi}$ contribute to the formation of the axial magnetic field in the magnetoquasistatic regime. Interestingly, according to Fig. (\ref{Fig4}) the amplitude of electromagnetic fields achieve higher values at lower frequencies when the ME effect become stronger. This feature is highlighted by the NI-TI and TI-TI configurations (notice that yellow and dark blue regions representing the largest positive and negative amplitudes move to higher frequencies as $\chi$ decreases), suggesting that the ME coupling could be employed for the generation of intensive electromagnetic fields at will. Nonetheless, this feature turns around for arbitrarily large values of both exciting frequencies and  ME polarizabilities: upon further inspection, it is also appreciated from Fig. (\ref{Fig4}) that the electromagnetic fields amplitude drop when $\chi$ and $\omega_{0}$ rise simultaneously (i.e. $r_{2}\omega_{0}/c_{3}> 3$ and $\chi> 500\alpha_{0}$). This will be more clear in the subsequent discussion for the electromagnetic energy. Despite what one could expect, by considering highly-alternating currents, we find that an intense ME coupling makes harder to produce electromagnetic fields with identical amplitude. In Sec. \ref{SecHFR}, we provide analytical results in the limit of large frequencies ($r_{2}\omega_{0}/c_{i}\gg 1$ with $i=1,2,3$) and strong ME responses ($r_{2}\mu_{0}\omega_{0}\alpha_{0}\gg 1$) that confirms this observation: a significantly strong ME response plays a degrading effect in the formation of highly-energetic electromagnetic fields (see Eqs. from \eqref{BLFAPPS} to \eqref{ELFAP1ZS}) that imposes a cutoff frequency on the ME solenoid inductor. 

In summary, we have shown that the influence of the ME effect upon the electromagnetic fields becomes significantly appreciable in the frequency domain of current power electronics when dealing with large values of the ME susceptibility in comparison with time-reversal-symmetry TIs ans AIs, i.e. $\chi\sim 1000\alpha_{0}$. In particular, the amplitude of the axial component $B_z$ significantly grows by increasing $\chi$ in the magnetoquasistatic regime, which is attributed to the rising of the azimuthal surface Hall currents. The latter suggests that the induction properties of the solenoid can be also modified as well by the presence of the ME effect. This question is extensively examine in the following section.

\begin{figure*}
\centering
\includegraphics[scale=0.42]{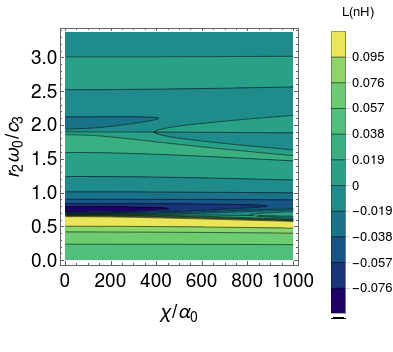}
\includegraphics[scale=0.42]{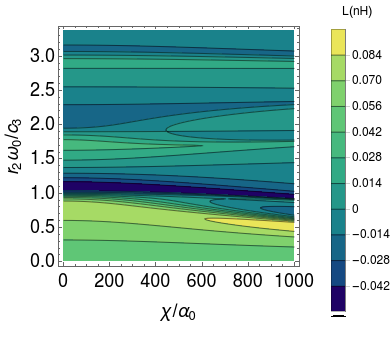}
\includegraphics[scale=0.42]{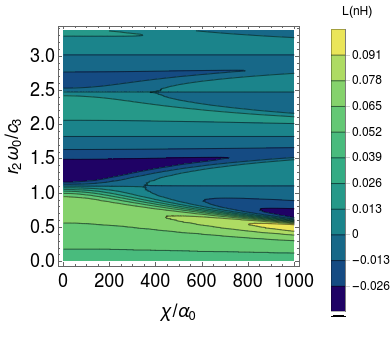}
\includegraphics[scale=0.42]{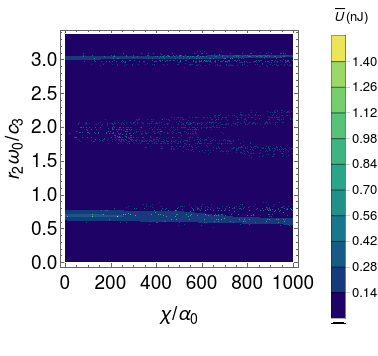}
\includegraphics[scale=0.42]{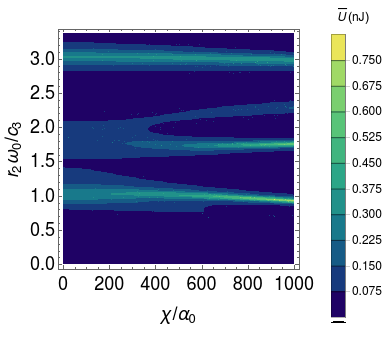}
\includegraphics[scale=0.42]{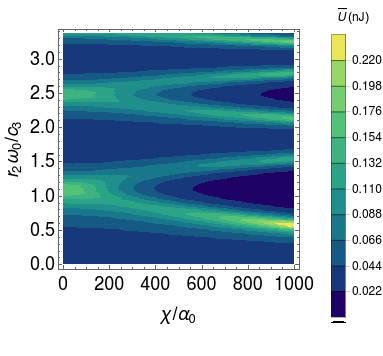}
\caption{(color online). The upper row contains contour plots of the self-induction coefficient as a function of the exciting frequency and the ME susceptibility for bilayer solenoid inductors sketched in Fig. 
 (\ref{Fig1Setup}): TI-NI (upper-left), NI-TI (upper-central), TI-TI (upper-right). The lower row also contains contour plots of the time-averaged electromagnetic energy as a function of the exciting frequency and the ME susceptibility: TI-NI (down-left), NI-TI (down-central), TI-TI (down-right). We have fixed the rest of parameters as in Fig. (\ref{Fig4}).  \label{Fig5}} 
\end{figure*} 
\subsection{Numerical computation of the self-induction coefficient and the electromagnetic energy} \label{Sec_NFSICEE}

We now compute the self-induction coefficient or inductance, which we shall denote by $L$, associated to the solenoid inductor by appealing to Eq. (\ref{Eqfemlow}). Concretely, we replace the magnetic flux through the solenoid cross-section returned by the axial magnetic field (\ref{BxBf}), and then we identify $L$ as the term accompanying the free circumferential current $I_{f}(t)$. This procedure retrieves the result expected for the self-induction coefficient of the conventional solenoid inductor \cite{harmon19911,thevenet19991,mcdonald19971,zangwill20121}. 

Fig. (\ref{Fig5}) shows the inductance in terms of the exciting frequency and the ME susceptibility for the studied bilayer configurations (see upper panels). For comparatively small values of $r_{2}\omega_{0}/c_{3}\sim 1/2$ (that is $\omega_{0}\sim 0.1$ THz), it can be seen that $L$ grows with the increase of $\chi$: notice that the self-induction coefficient gets higher values at lower frequencies specially in both the NI-TI and TI-TI configurations. As previously anticipated, this is because an additional axial component of the magnetic field arises from the azimuthal surface Hall currents (\ref{DCHSIP1}) and (\ref{DCHSIP2}). Notably, given a ME susceptibility $\chi\sim 1000\alpha_{0}$, the solenoid inductor composed of either the NI-TI or the TI-TI configuration manifests a self-inductance tunability of over $200\%$ up to 100 GHz in the millimeter length scale (which corresponds to the yellow region in the middle and right upper panels of Fig. (\ref{Fig5})). This is fairly comparable to previous proposals of ME voltage tunable inductors consisting of Metglas/PZT  and nickel/cobalt ferrite  composites, which have retrieved an inductance tunability around $50\%$ \cite{liang20211,lou20091,he20211} and $750\%$ \cite{yan20181} up to $10$ MHz, respectively (see also \cite{su20161,yan20201}).

It turns out that the main consequence of the ME response in the solenoid can be regarded as an additional inductance connected in series with the conventional setup in the magnetoquasistatic regime, we shall discuss this point in more detail in Sec.\ref{SecLFR}.  Nonetheless, the latter requires a significantly large ME response compared with standard TIs (i.e. $\chi\sim 1000\alpha_{0}$) since the ME effect represents a second-order perturbation to $L$ in the magnetoquasitatic regime (see Eq.(\ref{EqSelfindLowME})). It can be also noticed that $L$ is essentially uniform for changing $\chi$ when $\omega_{0}$ remains close to zero, which reflects the dynamical origin of the ME response once again. 

By paying attention to the limit of large values of both $\omega_{0}$ and $\chi$ in Fig. (\ref{Fig5}), one can appreciate that the (absolute) amplitude of $L$ monotonically decrease with increasing frequency, so that the ME effect eventually degrades the inductance. This is confirmed by our analytical study in Sec.\ref{SecHFR}: we find that $L$ become arbitrarily small for sufficiently large ME responses (i.e. $r_{2}\mu_{0}\omega_{0}\chi\gg 1$) in the high frequency domain. As a consequence, the operating frequency range could be substantially limited by the ME effect: for instance, it can be seen from Fig. (\ref{Fig5}) that for ME materials with $\chi\sim 1000\alpha_{0}$ the self induction is significantly suppressed beyond the RF domain (this corresponds to the upper right corner in the upper panels). The latter is consistent with the fact that the axial magnetic field may be largely diminished by strong ME effects at a given current frequency, previously unveiled by Fig. (\ref{Fig4}). We also emphasize that $L$ essentially carries the sign of the azimuthal magnetic field, so that $L$ may take on negative values. Actually, we shall show that the self-induction coefficient defined by (\ref{Eqfemlow}) represent a harmonic function of $\omega_{0}$ in the limit of high current frequencies and strong ME responses.

Additionally, we examine the time-averaged electromagnetic energy contained in the volume of the solenoid inductor, which gives an estimation of the energy consumption. This is given by 
\begin{equation}
    \overline{U}=\frac{\omega_0}{2\pi}
    \int_{0}^{2\pi/\omega_0}
    U(t) \, dt,
    \label{EUEMA} 
\end{equation}
once substituted Eqs. (\ref{ExEf}) and (\ref{BxBf}). Figure (\ref{Fig5}) illustrates $\overline{U}$ versus the ME susceptibility and the exciting frequency for the three studied configurations.  By comparing the results obtained for both $L$ (upper panels) and $\overline{U}$ (lower panels), one may observe that their behaviour is similar in terms of $\chi$ and $\omega_{0}$: the ME response is mainly manifested in the NI-TI and TI-TI configurations, while for the TI-NI scenario the changes of $\overline{U}$ could be essentially attributed to electric displacement effects. As anticipated in the previous section, by focusing on middle values of the current frequency (i.e. $r_{2}\omega_{0}/c_{3}\sim 1$), we may see that the highest values of the electromagnetic energy are achieved at lower frequencies when $\chi$ rises (notice that the yellow region moves down when we goes from the left to the right). This feature is due to the initial growth of the electromagnetic field amplitude, as stated before. However, the overall trend is inverted in the domain of arbitrarily large frequencies and ME coupling strengths: $\overline{U}$ apparently decreases as we consider higher values of $\chi$, which coincides with the discussion of Sec. \ref{Sec_NCEMF}. Based on the fact that the energetic cost of producing alternating currents is roughly characterized by $\omega_{0}$ \cite{zangwill20121}, these results mean that such cost initially diminishes for low and middle values of the ME coupling strength (i.e. $\chi< 500 \alpha_{0}$), though it eventually grows for sufficiently large values of the ME susceptibility (i.e.$\chi\gg \alpha_{0}$).

It can be concluded that the ME effect gives rise to substantial changes in both the solenoid self-inductance and time-averaged electromagnetic energy for comparatively small values of the frequency (including the RF domain) and relatively large ME couplings: for instance, the solenoid inductor composed of the NI-TI or the TI-TI configuration displays a self-inductance tunability of over $200\%$ up to 100 GHz for a ME susceptibility $\chi\sim 1000\alpha_{0}$ (and $r_{2}\sim 1$ mm). The latter value corresponds to conventional antiferromagnetic compounds rather than time-reversal TIs and AIs \cite{bhattacharyya20211,ying20221,ortega20151}). Additionally, we observe that a sufficiently strong ME response compared with the exciting frequency eventually diminishes the self-induction coefficient, which in turn causes a decrease of the operating frequency range: for example, this is restricted to the domain $1-10^{6}$ Hz in certain multiferroic compounds (recall that their ME susceptibility is around $\chi\sim 10^{3}\alpha_{0}$ and we are considering $r_{2}\sim 1$ mm). In order to confirm our numerical findings, next we present an extensive perturbative analysis of the electromagnetic quantities for interesting solenoid inductors. In particular, we shall explicitly show in the next section that $L$ can be viewed as two solenoid inductors connected in series: the first inductor corresponds to the self-induction coefficient of the conventional solenoid (in absence of the ME response), whereas the second inductor fully encodes the influence of the ME effect. We shall further show that for comparatively large values of the current frequency, the time-averaged electromagnetic energy gets arbitrarily small for sufficiently large ME coupling.

\subsection{Perturbative analysis}

In this section we provide an analytical study of electromagnetic fields, the self-induction coefficient, the time-averaged electromagnetic energy and time-averaged power radiated in two opposite scenarios: we first consider weak ME effects and low exciting frequencies, and then, we contrast these results with the case of strong ME effects and high exciting frequencies. For a clear exposition, we consider non-permeable media in this first part (i.e. $ \mu_{1,r}=\mu_{2,r}=1$).

\subsubsection{Weak ME response and low-frequency regime} \label{SecLFR}

Let us focus in the scenario of slowly varying currents and weak ME responses: that is, 
$r_{2}\omega_{0}/c_{i}\ll 1$ (for $i=1,2,3$) and $r_{2}\mu_{0}\omega_{0}\alpha_{0}\ll 1$. Since we are dealing with the low-frequency regime, it is convenient to assume $\omega_{0}\ll \omega_{R},\omega_{e}$; so that we can approximate the dielectric constant expression (\ref{EQDISRLF}).  Concretely, the subsequent analysis corresponds to the right-down corner in Figs. (\ref{Fig4}) and (\ref{Fig5}) when considering millimeter length scales. 

To derive expressions for the electromagnetic fields, we recall that Eqs. (\ref{BxBf}) and (\ref{ExEf}) represent complex combinations of the previously introduced Bessel and Hankel functions evaluated at $r$, $r_{1}$ y $r_{2}$. We first substitute their asymptotic forms for small arguments \cite{jackson20121}, and then, we perform a Taylor series expansion around $r_{2}\mu_{0}\omega_{0}\alpha_{0}=0$, $r_{2}\omega_{0}/c_{i}=0$, $r_{1}\omega_{0}/c_{i}=0$ and $r\omega_{0}/c_{i}=0$ (for $i=1,2,3$). After some manipulation, we get the following expressions of the magnetic field  components inside the first layer,
\begin{widetext}

\begin{align}
B_{\varphi}(r<r_{1},t)
&\approx  
\mu_{0}^2\frac{n I_{0}}{4\pi l}\alpha_{0}\omega_{0}
\Big[
r_{1}(\theta_1 - \theta_2) +r_{2}(\theta_2 - \theta_3)
\Big]
\sin(\omega_{0}t),\label{BLFAPP} \\
B_{z}(r<r_{1},t)
&\approx  
\mu_{0}\frac{n I_{0}}{4\pi l}
\Big[
4\pi+\frac{(\mu_{0}\alpha_{0}\omega_{0})^2}{\pi r_{2}}
\Big(
r_{1}^2r_{2}(\theta_{1}-\theta_{2})^2+(r_{1}^3+r_1r_2^2)(\theta_1-\theta_2)(\theta_2-\theta_3)+r_{2}^3(\theta_{2}-\theta_{3})^2
\Big) \nonumber \\
& \qquad \qquad 
+\pi\omega_{0}^2\Big(\frac{r_{1}^2}{c_{1}^2}+\frac{r_{2}^2-r_{1}^2}{c_{2}^2}\Big)+2\pi\Big(\frac{r_{2}\omega_{0}}{c_{3}}\Big)^2\Big(\log\Big(\frac{2c_{3}}{r_{2}\omega_{0}}\Big)-\gamma_{Euler}\Big)\Big]
\cos(\omega_0 t) \nonumber \\
& \quad +\pi\mu_{0}\frac{n I_{0}}{4l}\Big(\frac{r_{2}\omega_{0}}{c_{3}}\Big)^2\sin(\omega_{0}t),
    \label{BLFAPZ}
\end{align}
where $\gamma_{Euler}$ is the Euler's constant, and we have used the approximated expression for the light speed which derived from (\ref{EQPPERWN}), i.e.
\begin{equation}
    c_{lf}\approx\sqrt{\frac{c^2\omega_{R}^2}{\omega_{R}^2+\omega_{e}^2}},
    \label{EQDISLF} 
\end{equation}
such that $c_1=c_{lf}$ and $c_2=c_{lf}$ for the first and second arrangements, respectively; whereas $c_1=c_2=c_{hf}$ for the third configuration of the ME medium. Similarly, the electric field components inside the first layer read,
\begin{align}
E_{\varphi}(r<r_{1},t)
&\approx 
\mu_{0}\omega_{0}\frac{n I_{0}}{4\pi l}r
\Big[
2\pi 
+
\frac{(\mu_{0}\alpha_{0}\omega_{0})^2}{2\pi  r_{2}}
\Big(
r_{1}^2r_{2}(\theta_{1}-\theta_2)^2+(r_{1}^3+r_{1}r_{2}^2)(\theta_{1}-\theta_2)(\theta_{2}-\theta_{3})+r_{2}^3(\theta_{2}-\theta_{3})^ 2\Big)\nonumber \\
& \qquad \qquad \qquad +
\frac{\pi\omega_{0}^2}{2}\Big(\frac{r_{1}^2}{c_{1}^2}+\frac{r_{2}^2-r_{1}^2}{c_{2}^2}\Big)+\pi \Big(\frac{r_{2}\omega_{0}}{c_{3}}\Big)^2\Big(\log\Big(\frac{2c_{3}}{r_{2}\omega_{0}}\Big)-\gamma_{Euler}\Big)\Big]\sin(\omega_{0}t) \nonumber \\
& \quad -
\pi\mu_{0}\omega_{0}r\frac{nI_{0}}{8l}\Big(\frac{r_{2}\omega_{0}}{c_{3}}\Big)^2\cos(\omega_{0}t),  \label{ELFAPP} \\
E_{z}(r<r_{1},t)
&\approx \mu_{0}^2\frac{n I_{0}}{4\pi l}r\alpha_{0}\omega_{0}^2
\Big[
r_{1}(\theta_1 - \theta_2) +r_{2}(\theta_2 - \theta_3)
\Big]
\cos(\omega_{0}t),
    \label{ELFAPZ}
\end{align}
\end{widetext}
which clearly return the electromagnetic fields of the long vacuum solenoid excited by a direct current in the magnetostatic limit \cite{jackson20121}. Similarly, one may check that we recover the conventional solution of a vacuum long solenoid excited by an alternating current \cite{abbott19851,zangwill20121,harmon19911,thevenet19991,templin19951} after neglecting the ME effect (i.e. $\theta_{i}\rightarrow 0$ with $i=1,2$): in this case, both azimuthal $B_{\varphi}$ and longitudinal $E_{z}$ components cancels, as expected from the previous discussion. In App \ref{AppBLSSol}, we also provide the simplified expressions for the electromagnetic fields in the middle $r_{1}<r<r_{2}$ and outer $r_{2}<r $ regions  (see Eqs. from (\ref{BLFAPZ2a}) to (\ref{ELFAZbou})).

From Eqs. (\ref{BLFAPP})-(\ref{ELFAPZ}) it is clear that the ME effect constitutes a first-order correction to the azimuthal $B_{\varphi}$ and longitudinal $E_{z}$ components, while $B_{z}$ and $E_{\varphi}$ contain second-order corrections. These results explain the observation in Fig. (\ref{Fig3}): that is, the ME coupling remains hidden by the current displacement effects in the latter. Notice that the ME corrections depend explicitly on the arrangement of the TI layers: for instance, the first-order corrections to $B_{\varphi}$ and $E_{z}$ become comparatively larger in the scenario TI-TI, while the second-order corrections to $E_{\varphi}$ and $B_{z}$ take on larger values in the scenario NI-TI. 

By paying attention to Eqs. (\ref{BLFAPZ}) and (\ref{BLFAPZ2b}) once replaced (\ref{ELFAPP}) into expressions (\ref{DCHSIP1}) and (\ref{DCHSIP2}), the longitudinal magnetic field component in the inner region can be cast into the following form,
\begin{align}
B_{z}(r<r_{1},t)
&=\frac{ \mu_{0}}{l}
\Big(n I_{0}(t)+I_{\text{Hall},1}^{\varphi}(t)+I_{\text{Hall},2}^{\varphi}(t)\Big) \nonumber \\
&\quad +
\mathcal{O}\Big( 
\Big(\frac{r_{2}\omega_{0}}{c_{i}}\Big)^2+(r_{2}\mu_{0}\alpha_{0}\omega_{0})^3
\Big), \label{BzMQAa}
\end{align}
while it takes the form in the middle region,
\begin{align}
&    B_{z}(r_{1}<r<r_{2},t)
=
\frac{\mu_{0}}{l}
\Big(
n I_{0}(t)+I_{\text{Hall},2}^{\varphi}(t)\Big) \nonumber \\
&\qquad \qquad \qquad \quad +
\mathcal{O}\Big(
\Big(\frac{r_{2}\omega_{0}}{c_{i}}\Big)^2+(r_{2}\mu_{0}\alpha_{0}\omega_{0})^3
\Big), \label{BzMQAb}
\end{align}
which explicitly manifests the influence of the surface Hall currents. Remarkably, Eqs. (\ref{BzMQAa}) and (\ref{BzMQAb}) support the validity of the magnetoquasistatic approximation to compute the axial magnetic field: the ME influence is captured at leading order by the standard Ampère's law taking account the azimuthal surface Hall current as we anticipated in Sec. \ref{Sec_MRMC}.  Since the change of the piecewise magnetic relative permeabilities $\mu_{i,r}$ does not introduce substantial modifications in the Maxwell's equations (\ref{EoMAP}) and (\ref{EoMAZ}), one may expect that this result remains valid for a permeable ME medium. This point is further discussed in Sec. \ref{Sec_MCC}, where we also study the magnetic case of the TI-TI configuration with $\mu_{1,r}=\mu_{2,r}$ (see Eqs. (\ref{BLFAPP_M}) and (\ref{BLFAPZ_M}) in App. \ref{AppBLSSol}), and recover the solution of $B_{z}$ expected from the extended Ampère's law (\ref{MEQS4}) as well. Based on these evidences, we argue the validity of the latter to perform a magnetic circuit analysis in presence of the ME response. Concretely, regarding the treatment of Sec. \ref{Sec_MRMC}, we work within the parameter domain where expressions (\ref{BzMQAa}) and (\ref{BzMQAb}) hold (i.e. $r_{2}\omega_{0}/c_{i}\gg1$ for $i=1,2,3$ and $r_{2}\mu_{0}\omega_{0}\alpha_{0}\ll 1$), and
therefore, Eq. (\ref{MEQS4}) holds too. Upon considering the well-known prescriptions of magnetic circuits \cite{zangwill20121}, the latter leads to the extended Hopkinson's law (\ref{HopEq}). Let us emphasise that Eqs. (\ref{BzMQAa}) and (\ref{BzMQAb}) constitute a main finding of the present work, and have not been reported before to the best of our knowledge.

In the case of the azimuthal magnetic field, by substituting (\ref{DCHSZ1}) and (\ref{DCHSZ2}) in (\ref{BLFAPP}) and (\ref{BLFAPZ2a}), we obtain
\begin{align}
B_{\varphi}(r<r_{1},t)
&=\frac{\mu_{0}}{4\pi^2}
\Big(
\frac{I_{\text{Hall},1}^{z}(t)}{r_1}+\frac{I_{\text{Hall},2}^{z}(t)}{r_2}
\Big) \nonumber \\
& \quad +\mathcal{O}
\Big( \Big(\frac{r_{2}\omega_{0}}{c_{i}}\Big)^2+(r_{2}\mu_{0}\alpha_{0}\omega_{0})^4\Big), \label{BpMQA1}
\end{align}
and 
\begin{align}
& B_{\varphi}(r_{1}<r<r_{2},t)
=\frac{\mu_{0}}{4\pi^2}
\Big(
\frac{r_{1}I_{\text{Hall},1}^{z}(t)}{ r^2}+\frac{I_{\text{Hall},2}^{z}(t)}{r_2}
\Big) \nonumber \\
& \qquad \qquad \qquad  +
\mathcal{O}
\Big( 
\Big(\frac{r_{2}\omega_{0}}{c_{i}}\Big)^2+(r_{2}\mu_{0}\alpha_{0}\omega_{0})^4
\Big), \label{BpMQA2}
\end{align}
which differ from the well-known results returned by the Ampère's law for a straight line current \cite{jackson20121,zangwill20121} (recall this solution would be given by $B_{\varphi}= \frac{\mu_{0}I}{2\pi r}$). Unlike to the axial magnetic component, the Ampère's law fails to retrieve the correct azimuthal magnetic component. This proves that the Ampère's law is not trivially satisfied in the ME materials in the magnetoquasitatic regime.

To obtain the approximate self-induction coefficient, we compute the electromotive force via $\epsilon=-\frac{d\Phi_{m}}{dt}$ once replaced the magnetic flux $\Phi_{m}$ across the coil's section returned by the simplified expressions (\ref{BLFAPZ}) and (\ref{BLFAPZ2b}) for the magnetic field inside the solenoid. By casting the result in the form of Eq. (\ref{Eqfemlow}), we may directly identify the self-induction coefficient. We find that the latter contains an additional term, denoted by $L_{ME}$, that is
\begin{equation}
L
=
L_{0}
+L_{ME}
+\mathcal{O}
\Big( \Big(\frac{r_{2}\omega_{0}}{c_{i}}\Big)^4+(r_{2}\mu_{0}\alpha_{0}\omega_{0})^4\Big),
    \label{EqSelfindLowT}
\end{equation}
with $L_{0}$ being the self-induction coefficient of the vacuum solenoid, i.e.
\begin{align*}
L_{0}
&=\frac{\pi\mu_{0}n^{2}}{4l}
\Big[
4r_{2}^2+r_{1}^2\Big(\frac{r_1\omega_0}{c_{1}}\Big)^2 
+
r_{2}^2
\Big(1-\Big(\frac{r_{1}}{r_{2}}\Big)^4\Big)
\Big(\frac{r_2\omega_0}{c_{2}}\Big)^2\nonumber \\
& \qquad \qquad +
2r_{2}^2\Big(\frac{r_2\omega_0}{c_{3}}\Big)^2\Big(\log\Big(\frac{2c_{3}}{r_{2}\omega_{0}}\Big)-\gamma_{Euler}\Big)
\Big]
\end{align*}
and 
\begin{align} \label{EqSelfindLowME}
L_{ME}
&=
\frac{\mu_{0}^3n^2}{4\pi l}(\alpha_{0}\omega_{0})^2\Big[r_1^4(\theta_1-\theta_2)^2+r_2^4(\theta_2-\theta_3)^2  \nonumber \\
&\qquad \qquad \qquad \quad +
2r_1^3r_2(\theta_1-\theta_2)(\theta_2-\theta_3)\Big],
\end{align}
which fully contains the ME effect. The form of Eq. (\ref{EqSelfindLowT}) suggests that the leading-order ME response of the solenoid can be interpreted as an additional self induction $L_{ME}$ connected in series with $L_{0}$, as anticipated before. Alternatively, combining Eqs. (\ref{BzMQAa}) and (\ref{BzMQAb}) with the magnetic flux definition yields to the expression (\ref{EqSelfindLowT}), so this result works in the domain where we perform the magnetic circuit analysis in magnetic ME media as well. Notice that the influence of the ME response on the self induction explicitly depends on the bilayer configuration, indicating that we could slightly tune the magnetic flux. For instance, the NI-TI arrangement retrieves the highest contribution to the induction properties for given values of $r_{1}\neq r_{2}$, which is consistent with the fact that this configuration displays surface Hall currents flowing in both the inner as well outer cylindrical surfaces.  

Similarly, we find that the radiative term (which vanishes in the magnetostatic regime) takes the form
\begin{equation}
     N_{\text{rad}} \approx\frac{\mu_{0}\pi^2r_{1}^2 n^2 }{2lc}\Big(\frac{r_{2}\omega_{0}}{c_{3}}\Big)^2,
     \label{NradCoef}
\end{equation}
revealing that the radiative effects are independent of the ME response at leading order. In other words, the second term in the right hand side of (\ref{Eqfemlow}) essentially arises from the displacement electric effects, and thus, can be neglected in our analysis of the magnetic circuit by wisely choosing the current frequency  satisfying the magnetoquasistatic condition (\ref{DEEC}). By combining this result with Eq.(\ref{MEQS4}), we arrive to several expressions that represent the counterpart of the power converter equations \cite{zahn19791,wagemakers20171} in presence of a weak ME susceptibility. 

Additionally, from Eq. (\ref{EUEMA}) we obtain an approximate expression of the time-averaged electromagnetic energy after replacing from (\ref{BLFAPP}) to (\ref{ELFAPZ}) into (\ref{EUEM}). By doing some considerable manipulation, we get
\begin{widetext}
\begin{align}
&\overline{U}
\approx 
\frac{\pi\mu_{0}}{2l}(nI_{0})^2
\Big\{
r_{2}^2
+
\frac{5r_{1}^2}{8}\Big(\frac{r_{1}\omega_{0}}{c_{1}}\Big)^2
+
\frac{5r_{2}^2}{8}\Big(1-\Big(\frac{r_{1}}{r_{2}}\Big)^4 \Big)\Big(\frac{r_{2}\omega_{0}}{c_{2}}\Big)^2
+
r_{2}^2\Big(\frac{r_{2}\omega_{0}}{c_{3}}\Big)^2
\Big[
\Big(1-\Big(\frac{r_{1}}{r_{2}}\Big)^2 \Big)\log\Big(\frac{2c_{3}}{r_{2}\omega_{0}}\Big)-\gamma_{Euler}
\Big] \nonumber \\
&+
\Big(\frac{\mu_{0}\alpha_{0}\omega_{0}}{4\pi r_2}\Big)^2
\Big[
r_{1}^2
\Big(10(r_{1}r_{2})^2-r_{1}^4\Big) 
(\theta_{1}-\theta_{2})^2+9r_{2}^6 (\theta_{2}-\theta_{3})^2 +2(r_{1}r_{2})^3\Big(9+2\log\Big( \frac{r_{1}}{r_{2}}\Big)\Big) \left(\theta_{1}-\theta_{2}\right)\left(\theta_{2}-\theta_{3}\right) \Big]
\Big\}, \label{EEMELS}
\end{align}
where the first term in the right-hand side is the well-known electromagnetic energy of the long vacuum solenoid in the magnetostatic limit \cite{zangwill20121}. Interestingly enough, the ME effect represents a second-order correction to the electromagnetic energy as well as the displacement electric effect. Expression (\ref{EEMELS}) also tells us in the magnetoquasistatic regime that the energetic cost to create electromagnetic fields inside the solenoid grows as a quadratic power of the current frequency for a given value of the ME polarizability. 

Lastly, we analyse the radiation field generated by the bilayer long solenoid. We compute the time average of the power radiated per unit length (passing out through a cylinder of unit length over a period) \cite{zangwill20121},
\begin{equation}
\overline{P}
=
\lim_{r\rightarrow\infty} 
\omega_{0} r
\int_{0}^{2\pi/\omega_0}|\vect S(r,t)| \, dt.
   \label{PWRC}
\end{equation}
After integrating once substituted the simplified expression for the electromagnetic fields (see equations from (\ref{BLFAPa}) to (\ref{ELFAZbou})), we find
\begin{align}
\overline{P} 
&\approx2\mu_{0}\omega_{0}\Big(\frac{\pi n I_{0}}{4l}\Big)^2\Big(\frac{r_{2}\omega_{0}}{c_{3}}\Big)^2
\Big\{
r_{2}^2 +\frac{r_{1}^2}{2}\Big(\frac{r_{1}\omega_{0}}{c_{1}}\Big)^2 +\frac{r_{2}^2}{2}\Big(1-\Big(\frac{r_{1}}{r_{2}}\Big)^4 \Big)\Big(\frac{r_{2}\omega_{0}}{c_{2}}\Big)^2    \nonumber \\
& \quad +
\frac{(\mu_{0}\alpha_{0}\omega_{0})^2}{4\pi^2}\big[( r_1^6+2r_{1}^4r_{2}^2)\left(\theta_{2}-\theta_{1}\right)^2+6(r_1r_2)^3 \left(\theta_{1}-\theta_{2}\right)\left(\theta_{2}-\theta_{3}\right)+3r_{2}^6\left(\theta_{2}-\theta_{3}\right)^2\big]
\Big\}, \label{PradCoef}
\end{align}
\end{widetext}
where the first term in the right-hand side coincides with the time-averaged radiation power of the long vacuum solenoid excited by an alternating current \cite{abbott19851,zangwill20121}, while the second term completely encodes the radiative effect due to a weak ME response. One may appreciate that the NI-TI configuration provides the highest radiative effects, which is expected from the preceding discussion.

\subsubsection{Strong ME response and high-frequency regime}\label{SecHFR}

Let us now turn the attention on the opposite scenario, when we deal with high frequencies and strong ME susceptibilities, we thus adhere to $r_{2}\omega_{0}/c_{i}\gg 1$ (for $i=1,2,3$) and $r_{2}\mu_{0}\omega_{0}\alpha_{0}\gg 1$. With respect to Figs. (\ref{Fig4}) and (\ref{Fig5}), this analysis corresponds to the right-upper corner.  Since we are dealing with the high-frequency regime (i.e. $\omega_{0}\gg \omega_{R}$), it is convenient to make use of the reduced expression (\ref{EQDISRHF}) of the dielectric constant.

Here we perform a Taylor series expansion up to first order in both perturbative parameters $(r_{2}\omega_{0}/c_{i})^{-1}$, $(r_{1}\omega_{0}/c_{i})^{-1}$ and $(r_{2}\mu_{0}\omega_{0}\alpha_{0})^{-1}$, after replacing the asymptotic expression of the Bessel and Hankel functions for large arguments \cite{jackson20121}. We get for the magnetic field components,
\begin{widetext}

\begin{align}
    B_{\varphi}(r<r_{1},t)&\approx  \frac{\sqrt{r_{2}\pi^9k_1^3}k_2nI_{0}\csc(k_2(r_1-r_2))(J_{0}(k_1 r)-J_{2}(k_1 r))}{2\pi l\mu_{0}^2(\alpha_{0}\omega_{0})^3\left(\theta_{1}-\theta_{2}\right)^2\left(\theta_{2}-\theta_{3}\right)^2(\sin(2k_1r_{1})-1)}
    \Big[-k_3\left(\theta_{1}-\theta_{2}\right)(\cos(k_1r_{1})-\sin(k_1r_{1}))\cos(\omega_{0}t) \nonumber \\
    & \quad +(\cos(k_1r_1)u_{+}(k_1,k_2)+\sin(k_1r_1)u_{-}(k_1,k_2))\sin(\omega_{0}t)\Big], \label{BLFAPPS} \\
     B_{z}(r<r_{1},t)&\approx  \frac{\sqrt{r_{2}\pi^7k_1^3}k_2n I_{0}\csc(k_2(r_1-r_2))J_{0}(k_1 r)\cos(\omega_{0}t)}{\pi l\mu_{0}(\alpha_{0}\omega_{0})^2\left(\theta_{1}-\theta_{2}\right)\left(\theta_{2}-\theta_{3}\right)(\cos(k_1r_{1})-\sin(k_1r_{1}))},   \label{BLFAPZS}
\end{align}
and the electric field components,
\begin{align}
     E_{\varphi}(r<r_{1},t)&\approx \frac{\sqrt{r_{2}\pi^7k_1}k_2nI_{0}\csc(k_2(r_1-r_2))J_{1}(k_1 r)\sin(\omega_{0}t)}{\pi l\mu_{0}\alpha_{0}^2\omega_{0}\left(\theta_{1}-\theta_{2}\right)\left(\theta_{2}-\theta_{3}\right)(\cos(k_1r_{1})-\sin(k_1r_{1}))}, \label{ELFAPPS} \\
E_{z}(r<r_{1},t)
&\approx \frac{\sqrt{r_{2}\pi^9k_1}k_2 nI_{0}\csc(k_2(r_1-r_2))J_{1}(k_1 r)}{ \pi l (\mu_{0}\omega_{0})^2\alpha_{0}^3\left(\theta_{1}-\theta_{2}\right)^2\left(\theta_{2}-\theta_{3}\right)^2(\sin(2k_1r_{1})-1)}
\Big[k_3\left(\theta_{1}-\theta_{2}\right)(\cos(k_1 r_{1})-\sin(k_1r_{1}))\sin(\omega_{0}t) \nonumber  \\
     & \quad +(\cos(k_1r_1)u_{+}(k_1,k_2)+\sin(k_1r_1)u_{-}(k_1,k_2))\cos(\omega_{0}t)
\Big], \label{ELFAP1ZS}
\end{align}
where we have introduced the auxiliary function 
\begin{equation*}
u_{\pm}(k_1,k_2)
=k_1\left(\theta_{2}-\theta_{3}\right)\pm k_2\left(\theta_{1}-\theta_{3}\right)\cot(k_2(r_{1}-r_{2})).
\end{equation*} 
and employed the asymptotic expression of the dispersion relation for the ME material (obtained from Eq. (\ref{EQPPERWN}) after replacing (\ref{EQDISRHF})), i.e. 
\begin{equation}
    k_{hf}(\omega_{0})=\frac{\omega_{0}}{c}\sqrt{1-\Big(\frac{\omega_{e}}{\omega_{0}}\Big)^2},
    \label{EQDISHF}
\end{equation}
Notice that $k_1=k_{hf}(\omega_{0})$ and $k_2=k_{hf}(\omega_{0})$ for the first and second arrangements, respectively; whereas $k_1=k_2=k_{hf}(\omega_{0})$ for the third configuration of the ME medium. Interestingly enough, Eqs. from (\ref{BLFAPPS}) to (\ref{ELFAP1ZS}) reveal a degrading effect of the ME response upon the electromagnetic fields: these become arbitrarily small by sufficiently large ME susceptibility for a fixed exciting frequency. This result was anticipated by Fig. (\ref{Fig4}) and can be alternatively viewed as follows: higher values of $\omega_{0}$ are required to produce electromagnetic fields with the same desirable amplitude as the ME polarizability becomes significantly large. The latter implies that the generation of highly-energetic electromagnetic fields becomes energetically expensive in the presence of strong ME effects. This point will become clear below once we compute the solenoid electromagnetic energy: it turns to be a decreasing power of the ME coupling coefficient. From Eqs. (\ref{BLFAPPS}) to (\ref{ELFAP1ZS}), one may also appreciate that the geometry of the bilayer configuration ultimately
characterizes the electromagnetic fields, as similarly occurs for low frequencies and weak ME coupling strengths.

We further study the induction coefficients, as well as the time average of the electromagnetic energy and power radiated. These have rather tedious expressions in the strong ME scenario for generic setups, here we only discuss the TI-TI arrangement (for which holds $r_{1}=r_{2}$, $c_1=c_2$, and $\theta_{1}=\theta_{2}$) for sake of clarity. Hence simplified expressions for the self-induction and radiative coefficients are obtained by repeating the procedure described in the previous section, except we now employ the azimuthal magnetic field (\ref{BLFAPPSTITI}). The latter returns
\begin{align}
L
\approx\frac{2\pi^2n^2\Big(\cos(k_{2}r_{2})+\sin(k_{2}r_{2})\Big)J_{1}(k_{2}r_{2})}{ l\mu_{0}(\omega_{0}\alpha_{0})^2(\theta_{2}-\theta_{3})^2(\sin(2k_{2}r_{2})-1)}\Big(\frac{r_{2}\omega_{0}}{c_{2}}\Big)^{3/2}, \label{LSelfInMEH}
\end{align}
and 
\begin{equation}
  N_{\text{rad}}\approx\frac{2\pi^2n^2 J_{1}(k_{2}r_{2})}{l\mu_{0}(\omega_{0}\alpha_{0})^2(\theta_{2}-\theta_{3})^2
  \Big(-\cos(k_{2}r_{2})+\sin(k_{2}r_{2})\Big)}\frac{(r_{2}\omega_{0})^{3/2}}{c_{3}\sqrt{c_{2}}}. \label{NSelfInMEH}
\end{equation}
which reveals that the ME effect make arbitrarily small the solenoid induction coefficients at a given value of the exciting frequency. As expected from previous discussion, the ME response has also a harmful effect upon the solenoid induction properties. By paying attention to Eq. (\ref{NSelfInMEH}), one may also understand the change of sign in $L$ for large values of frequencies: the self-induction coefficient probes to be a harmonic function in the current frequency since $k_{2}r_{2}\propto \omega_{0}$. The latter explain the fact that the self-induction coefficient in Fig. (\ref{Fig5}) may take on negative values, essentially because the sign of the $B_{z}$ changes with $\omega_{0}$.

Similarly, after doing a Taylor expansion once substituted the approximate electromagnetic fields (see Eqs. from (\ref{BLFAPPSTITI}) to (\ref{ELFAP1ZSTITI}) in App. \ref{AppBLSSol}) in (\ref{EUEM}) and (\ref{EUEMA}), we obtain the time-averaged energy of the electromagnetic energy stored by the solenoid in the TI-TI setup,
\begin{align}
\overline{U}
&\approx\frac{(\pi^2 nI_{0})^2}{2l\mu_{0}(\omega_{0}\alpha_{0})^2(\theta_{2}-\theta_{3})^2\Big(\sin(2k_{2}r_{2})-1\Big)}\Big(\frac{r_{2}\omega_{0}}{c_{2}}\Big) \nonumber \\
& \quad \times 
\Big[1 -J_{0}^2(k_{2}r_{2})+J_{1}^2(k_{2}r_{2})
 +2\Big(\frac{r_{2}\omega_{0}}{c_{2}}\Big)J_{0}(k_{2}r_{2})J_{1}(k_{2}r_{2})-2\Big(\frac{r_{2}\omega_{0}}{c_{2}}\Big)^2
\Big(J_{0}^2(k_{2}r_{2})+J_{1}^2(k_{2}r_{2})\Big)  
\Big], \label{UMEH}
\end{align}
\end{widetext}
which clearly diminishes with the growth of the ME coupling strength. As anticipated above, in order to produce a certain amount of electromagnetic energy, Eq. (\ref{UMEH}) predicts that higher frequencies are required for larger ME couplings. This contrasts with the conventional solenoid inductor in absence of the TME response. Upon further inspection of (\ref{UMEH}), one may realize that $\overline{U}$ grows for increasing values of the $\omega_{0}$ despite the degrading influence of the ME effects upon the electromagnetic fields (notice that the fifth term in the right hand side of Eq. (\ref{UMEH}) effectively goes as $\sim \omega_{0}/\alpha_{0}$). In other words, we need more energetically expensive currents (recall that the energetic cost of creating such currents is mainly determined by $\omega_{0}$ \cite{zangwill20121}) to produce the same electromagnetic energy into solenoids where the ME response becomes stronger. 

Finally, we provide the time average of the power radiated by the solenoid at leading order in the strong ME effects,
\begin{align*}
   \overline P  &\approx\frac{r_{2}(\pi nI_{0})^2}{2\mu_{0}l^2c_3\alpha_{0}^2(\theta_{2}-\theta_{3})^2},
\end{align*}
which unveils that the emitted radiation saturates at certain value independent of the current frequency. This contrasts with the conventional situation in absence of the ME effect, where $\overline P$ displays a linear growth with $\omega_{0}$ \cite{zangwill20121,thevenet19991,rivera20021}.

\begin{figure}[ht]
    \includegraphics[scale=0.4]{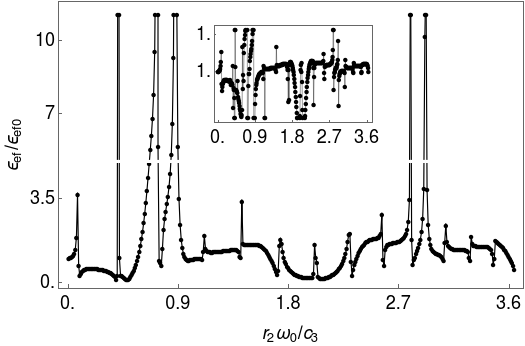}
    \includegraphics[scale=0.4]{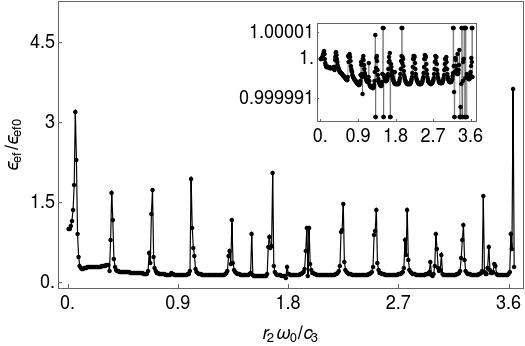}
    \includegraphics[scale=0.4]{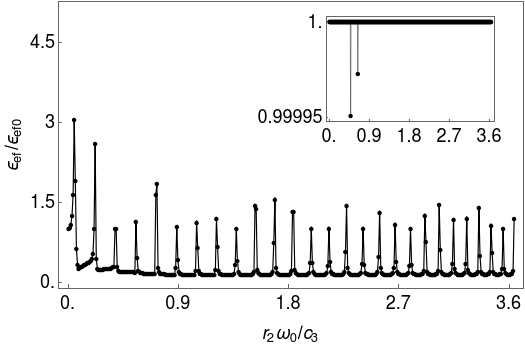}
\caption{Effective electromotive force induced in a ME primitive transformer for the three configurations of the core endowed with a strong ME susceptibility $1000\alpha_{0}$ and a high relative permeability equal to 100: TI-NI (upper), NI-TI (central), TI-TI (down). The inset depicts the effective electromotive force retrieved by a ME primitive transformer composed of a magnetic topological layer with an identical permeability and featuring the TME effect (notice that the only difference is $\chi=\alpha_{0}$). Here all the results are normalized to the effective electromotive force $\epsilon_{\text{ef0}}$ in absence of the ME response. The black dots signal the numerical results obtained by numerical integration, and the rest of parameters were fixed as in Fig. (\ref{Fig4}).\label{Fig6}} 
\end{figure} 

To recap our previous results, on one hand, we have shown that the Ampère's law applied to the axial magnetic component holds for comparatively low frequencies and weak ME effects: interestingly enough, we recover an extension of the standard results of $B_{z}$ in the magnetoquasistetatic regime that includes the azimuthal surface Hall current $I_{\text{Hall},i}^{\varphi}$ ($i=1,2$). This result supports the magnetic circuits analysis made in Sec. \ref{Sec_MRMC}. Furthermore, we identify a second-order correction, namely $L_{ME}$, to the self-induction coefficient that fully characterises the solenoid response against the ME effect in the magnetoquasistetatic domain. On the other hand, we have also shown that a large ME susceptibility make more energetically expensive to produce the same desirable amount of electromagnetic energy. Concretely, we found out that the ME response has a degrading effect upon the electromagnetic fields, such that the latter becomes arbitrarily small for sufficient large values of the former (if the current frequency is maintained fixed). As a consequence, the self-induction properties of the solenoid inductor become demeaned by a strong ME response, and thus, the latter is inconvenient for the construction of magnetic circuits.

\section{Magnetically coupled systems}\label{Sec_MCC}

The discussion of the preceding section mainly focuses on non-permeable media, however we shall show in this section that the theoretical treatment presented holds for magnetically coupled systems as well. Now we examine the impact of the ME effect upon isotropic, homogeneous magnetic media, such that $\mu_{1,r}$ and $\mu_{2,r}$ may take on arbitrary values, which comprise the situation of AIs and generic magnetic ME materials. Concretely, we study the instructive example of a magnetically coupled circuit, that is the anticipated primitive transformer \cite{jackson20121}. The latter is sketched in Fig. (\ref{Fig1Setup}.d) and basically consists of previous bilayer solenoid inductors with a second coil placed in the outer surface (with $r=r_{2}$), which is referred to as secondary winding (while the previous coil play the role of the primary winding). This allows to employ our previous results to further investigate the induced electromotive force $\epsilon_{\text{ef}}$ in the second coil as well as the magnetic loss of the ME primitive transformer, and compare them with the conventional situation in absence of the ME response. Finally, we close this section studying the case of another prominent example of magnetically coupled circuits: the solenoid actuator. Concretely, we compute the magnetic force arising from the ME effect at leading order in the weak ME susceptibility. Since most of the applications of these setups are exploited for low current frequencies \cite{apicella20191}, hereafter we focus the attention on the magnetoquasitatic regime. 

\subsection{Primitive transformer} 

As emphasized above, the magnetic core is shared by the coils in the primitive transformer, so that the magnetic flux $\Phi_{m}$ across the solenoid section is identical in both coils. We may determine the latter from the azimuthal component of the magnetic field (\ref{BxBf}) and then substitute the result in the integral expression of the Faraday's law (i.e. $\epsilon=-\frac{d\Phi_{m}}{dt}$) to obtain the induced electromotive force. Instead, since we are dealing with ac electromagnetic fields, it is convenient to analyse the effective electromotive force, i.e.  
\begin{equation}
\epsilon_{\text{ef}}
=\Big(\frac{\omega_0}{2\pi}
\int_{0}^{2\pi/\omega_0}\epsilon(t)^2 dt\Big)^{1/2}.
    \label{EEEMTF}
\end{equation}

Figure (\ref{Fig6}) illustrates the ratio of $\epsilon_{\text{ef}}$ to the effective electromotive force in absence of the ME response (which is denoted by $\epsilon_{\text{ef0}}$) versus the exciting frequency, given a high magnetic permeability of the TI, say $\mu$. We may see that for a weak ME response (recall that $\chi=\alpha_{0}$) and low frequencies, there is none appreciable change with respect to the conventional situation independently of the value of $\omega_{0}$ (see inset, notice that $\epsilon_{\text{ef}}/\epsilon_{\text{ef0}}$ is essentially constant and approximately equal to unit). That is, a weak ME response combined with a strong bulk magnetization does not retrieve further induction properties beyond the conventional situation (with identical bulk magnetization) despite the coupling between the electric and magnetic fields. Indeed, we shall show below that the ME response represents a second-order perturbation to $\epsilon_{\text{ef}}$ (see Eq. (\ref{EEMTFLSME1})) in the limit of high permeability (i.e.$\mu_{2,r}\gg 1$) as similarly found for the non-permeable case in Sec. \ref{SecLFR}. Moreover, the latter indicates that the results of Sec. \ref{SecLFR} holds for permeable materials as well.

Interestingly, in the opposite scenario of a strong ME response, the effective electromotive force displays a resonance-like pattern: $\epsilon_{\text{ef}}/\epsilon_{\text{ef0}}$ remains close to zero except for specific values of $\omega_{0}$, for which it takes on larger values compare to the conventional situation, specially in the TI-NI setup. The fact that $\epsilon_{\text{ef}}$ gets arbitrarily small values can be roughly understood by recalling that a strong ME response diminishes the amplitude of the electromagnetic fields at a constant frequency (see Eq. (\ref{BLFAPZS}) in Sec. \ref{SecHFR}). Although the structure of the peaks changes from one setup to another, by further inspection one may appreciate that the peak at lowest frequency is located approximately at the same value $\omega_{0}$ in the three configurations. Because of the complex nature of the ME coupling it is difficult to elucidate the condition for such resonance.

We now provide simplified expressions for the induced electromotive force generated in the secondary coil by electromagnetic induction in the TI-TI setup (recall $\theta_{1}=\theta_{2}$, $r_{1}=r_{2}$, $ \mu_{1,r}=\mu_{2,r}$). More specifically, by starting from an approximated solution of the azimuthal magnetic field (which is given by Eq. (\ref{BLFAPZ_M}) in App. \ref{AppBLSSol}), we obtain a close form expression of $\epsilon_{\text{ef}}$ for weak ME effects and low exciting frequencies (that is, $r_{2}\omega_{0}/c_{3} \ll 1$ and $r_{2}\mu_{0}\omega_{0}\alpha_{0} \ll 1$). After doing this, we further carry out a series expansion in $\mu_{2,r}\gg 1$ to get the induced electromotive force behaviour at leading order in the magnetic permeability, which yields
\begin{equation}
  \epsilon_{\text{ef}}=  \epsilon_{\text{ef0}}+\epsilon_{\text{ME}}+\mathcal{O}\bigg((r_{2}\mu_{0}\omega_{0}\alpha_{0})^4+\Big(\frac{r_{2}\omega_{0}}{c_{3}}\Big)^4\bigg),
\end{equation}
where $\epsilon_{\text{ME}}$ fully encodes the effective electromotive force exclusively arising from the ME response, that this
\begin{align}
\epsilon_{\text{ME}}
=\frac{r_2^2 \mu_{0}\omega_{0}n I_{0}}{2\pi\sqrt{2}l}(\theta_{2}-\theta_{3})^2\mu_{2,r}^2(r_{2}\omega_{0}\alpha_{0})^2.
    \label{EEMTFLSME1}
\end{align}
The interesting reader can also find a reduced expression of the effective electromagnetic force for non-permeable media in App. \ref{AppBLSSol}, see Eq. (\ref{EQEMTFV}). Equation (\ref{EEMTFLSME1}) reveals that the electromagnetic induction due to both the surface Hall current and the free current are identically enhanced by an intrinsic magnetization. This result can be intuitively understood by paying attention to the linearity of the ME coupling in the constitutive relation (\ref{ECT1}): it is clear that the magnetization induced by the ME response directly sums up to the conventional magnetization owing to the bound currents \cite{zangwill20121,jackson20121}. Let us notice that the axial magnetic field, which retrieves (\ref{EEMTFLSME1}), fulfills the modified Ampère's law (\ref{MEQS4}). This supports our results of Sec. \ref{Sec_MRMC} when they are applied to a highly permeable medium.

Additionally, for the TI-TI configuration, we study the induced electromotive force in the limit of strong ME effects but low frequencies and high magnetic permeability (that is, now $r_{2}\mu_{0}\omega_{0}\alpha_{0} \gg 1$ whereas $r_{2}\omega_{0}/c_{3} \ll 1$ and $\mu_{2,r}\gg 1$ remains as above). This corresponds to the situation represented in the main frame of the lower panel of Fig. (\ref{Fig6}). After some manipulation we obtain up to leading order
\begin{align}
\epsilon_{\text{ef}}
&\approx\frac{2\pi^{2}n I_{0}}{l\left(\theta_{2}-\theta_{3}\right)^2(\mu_{0}\omega_{0})^2\alpha_{0}^4}
\Big[1 + \frac{1}{\mu_{2,r}}
\Big(1-\frac{3}{8}\Big(\frac{r_{2}\omega_{0}}{c_{3}}\Big)^2\Big)\nonumber \\
& \quad - \Big(\frac{r_{2}\omega_{0}}{c_{3}}\Big)^2\Big(\log\Big(\frac{2c_{3}}{r_{2}\omega_{0}}\Big)-\gamma_{Euler}\Big)
\Big] \label{EEMFLFWME}
\end{align}
which reveals that $\epsilon_{\text{ef}}$ is eventually suppressed for a strong ME polarizability at a constant value of the exciting frequency. Remarkably enough, we also appreciate that, at a fixed $\omega_{0}$, $\epsilon_{\text{ef}}$ saturates to certain value independent of the magnetic permeability of the ME medium. This means that the conventional magnetization effects are significantly mitigated by the magnetization stemming on the ME response, which we called $\theta$-magnetization.

\begin{figure}[ht]
    \includegraphics[scale=0.40]{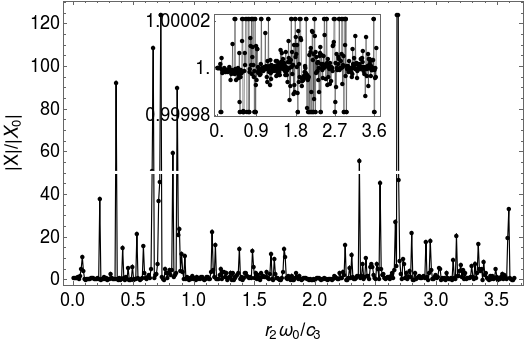}
    \includegraphics[scale=0.40]{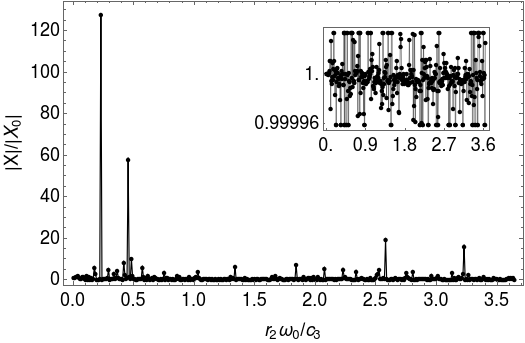}
    \includegraphics[scale=0.40]{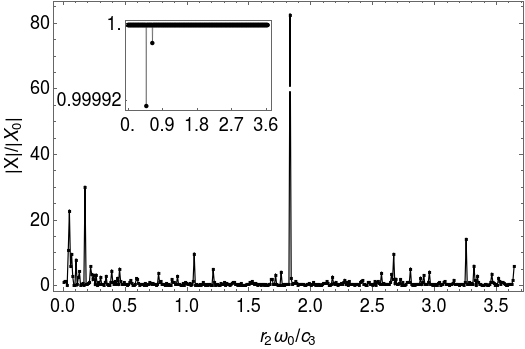}
\caption{The norm of the magnetic leakage reactance in a ME primitive transformer for the three configurations of the permeable core endowed with a strong ME susceptibility $1000\alpha_{0}$ and a high relative permeability equal to 100: TI-NI (upper), NI-TI (central), TI-TI (down). The inset depicts the magnetic leakage reactance retrieved by a ME primitive transformer composed of a magnetic topological layer with an identical permeability and featuring the TME effect (notice that the only difference is $\chi=\alpha_{0}$). Here all the results are normalized to the magnetic leakage reactance $X_{0}$ in absence of the ME response. The black dots signal the numerical results obtained by numerical integration, and the rest of parameters were fixed as in Fig. (\ref{Fig4}). \label{Fig7}} 
\end{figure}

\subsection{Real transformer}\label{Sec_RealT}
In practice, magnetically coupled systems would suffer from several issues, among them we highlight the magnetic flux produced by the primary winding that does not link the secondary winding: this is characterized by two leakage reactances $X^{(1)}$ and $X^{(2)}$ introduced in Sec. \ref{Sec_MRMC}. Both are identical in the primitive transformer, namely $X^{(1)}=X^{(2)}=X=i\omega_{0} L_{c}$, where $L_{c}$ correspond to the magnetic flux located in the free space region surrounding the coils. In other words, $L_{c}$ can be estimated by computing the amount of magnetic flux outside the solenoid~\cite{zahn19791}. In our case, this can be approximately computed by 
\begin{equation*}
L_{\text{c}}=\lim_{R\rightarrow\infty}2\pi\int_{r_{2}}^{R}\bigg( \frac{1}{r}\text{Re} \ \hat A_{\varphi}(r,\omega_{0})+\frac{\partial}{\partial r}\text{Re} \ \hat A_{\varphi}(r,\omega_{0}) \bigg) r dr,
    \label{LICT}
\end{equation*}
where $R$ is chosen arbitrarily large until getting a desirable convergence. We shall denote the magnetic leakage reactance in absence of ME response as $X_{0}$.

Figure (\ref{Fig7}) depicts the ratio of $|X|$ to $|X_{0}|$, as a function of the current frequency. Upon inspection of the inset, which corresponds to the weak ME susceptibility scenario (i.e $\chi=\alpha_{0}$), one may see that the magnetic leakage reactance in all studied setups largely coincides with the conventional situation in absence of the ME response: notice that $|X|/|X_{0}|$ remains close to unit. This suggests that the magnetic leakage exclusively arising from the ME response, say $X_{ME}$, represents, at most, a second-order perturbative correction in $r_{2}\mu_{0}\omega_{0}\alpha_{0} \ll 1$ to the conventional magnetic leakage, which is consistent with our discussion around the axial magnetic field in Sec. \ref{Sec_NCEMF}. Similarly as before, the behavior of the magnetic leakage substantially changes in the strong ME response domain: concretely,  we find a resonance-like pattern as well. Again, though it is not appreciated, the peak at lowest frequency is located approximately at the same value $\omega_{0}$ in the three configurations.

\begin{table*}[t!]

\begin{center}
\begin{tabular}{c|cc|cc|cc}
\hline
\hline
&\multicolumn{2}{c}{TI-NI} & \multicolumn{2}{c}{NI-TI} & \multicolumn{2}{c}{TI-TI}  \\  \cline{1-7}
$\omega_{0}$ &
$1-10^{12}$ Hz& $>10^{14}$ Hz&  $1-10^{12}$ Hz   & $>10^{14}$ Hz &  $1-10^{12}$ Hz   & $>10^{14}$ Hz \\
  $\chi$ & $ \alpha_{0}$   & $>10^6\alpha_{0}$ &  $ \alpha_{0}$   & $>10^6\alpha_{0}$ &  $ \alpha_{0}$   & $>10^6\alpha_{0}$ \\ 
\hline
$L$  & $L_{0}+\frac{\mu_{0}^3\pi(nr_{1}^2\alpha_{0}\omega_{0})^2}{4l}$ &  - &  $L_{0}+\frac{\mu_{0}^3\pi(n\alpha_{0}\omega_{0})^2h(r_{1},r_{2})}{4l}$ &- &  $L_{0}+\frac{\mu_{0}^3\pi(nr_{2}^2\alpha_{0}\omega_{0})^2}{4l}$ &  $\sim \sqrt{\frac{n^4 r_{2}^3}{(l\mu_{0}\alpha_{0}^2)^2\omega_{0}c_{2}^3}}$ \\
$N_{\text{rad}}$  & $\sim\big(\frac{r_{2}\omega_{0}}{c_{3}}\big)^2$ &  - &  $\sim\big(\frac{r_{2}\omega_{0}}{c_{3}}\big)^2$ & - &  $\sim\big(\frac{r_{2}\omega_{0}}{c_{3}}\big)^2$ &  $\sim \sqrt{\frac{n^4 r_{2}^3}{(l\mu_{0}c_{3}\alpha_{0}^2)^2\omega_{0}c_{2}}}$ \\ 
$\bar{U}$  & $\bar{U}_{0}+\frac{\pi\mu_{0}^3( r_{1}\omega_{0}\alpha_{0}n I_{0})^2}{32lr_{2}^2(10(r_{1}r_{2})^2-r_{1}^4)^{-1}}$ &  - &  $\bar{U}_{0}+\frac{\pi\mu_{0}^3(\omega_{0}\alpha_{0}n I_{0})^2p(r_{1},r_{2})}{32l r_{2}^2}$ & - &  $\bar{U}_{0}+\frac{9\pi\mu_{0}^3(\pi r_{2}^2\omega_{0}\alpha_{0}nI_{0})^2}{32l}$ &  $\sim \frac{r_{2}(nI_{0})^2}{l\mu_{0}\omega_{0}\alpha_{0}^2c_{2}}$ \\ 
$\bar{P}$  & $\bar{P}_{0}+\frac{\mu_{0}^3\omega_{0}^5(\pi r_{2}\alpha_{0}n I_{0})^2}{2c_{3}^2(r_{1}^6+2(r_{1}^2r_{2})^2)^{-1}}$ &  - & $\bar{P}_{0}+\frac{\mu_{0}^3\omega_{0}^5(\pi r_{2}\alpha_{0}n I_{0})^2q(r_{1},r_{2})}{2c_{3}^2}$  & - &  $\bar{P}_{0}+\frac{3\mu_{0}^3\omega_{0}^5(\pi r_{2}^4\alpha_{0}n I_{0})^2}{2c_{3}^2}$ & $\sim\frac{r_{2}(nI_{0})^2}{\mu_{0}l^2c_{3}\alpha_{0}^2}$ \\
$\frac{\epsilon_{ME}}{\epsilon_{\text{ef0}}}$  & $\sim(r_{2}\mu_{0}\omega_{0}\alpha_{0})^2$ &  - & $\sim(r_{2}\mu_{0}\omega_{0}\alpha_{0})^2$ & - &  $\sim\frac{\pi\mu_{2}^{2} nI_{0}\omega_{0}^3(r_{2}^2\alpha_{0})^2}{\sqrt{8}l\mu_{0}}$ &  - \\
$\frac{|X_{ME}|}{|X_{0}|}$  & $\sim(r_{2}\mu_{0}\omega_{0}\alpha_{0})^2$ & - &  $\sim(r_{2}\mu_{0}\omega_{0}\alpha_{0})^2$ & - &  $\sim(r_{2}\mu_{0}\omega_{0}\alpha_{0})^2$ &  - \\
\hline
\hline
\end{tabular}
    
\end{center}
 \vspace{ - 05 mm}
\caption{Summary of the approximate values of electromagnetic magnitudes for the studied solenoid inductors at leading order in the low (see the first, third and fifth columns) and high (see the second, fourth and sixth columns) frequency domain, as well as for weak (see the first, third and fifth columns) and strong (see the second, fourth and sixth columns) ME susceptibilities in the millimeter length scale. $L_{0}$, $N_{\text{rad}0}$, $\bar{U}_{0}$, $\bar{P}_{0}$, $\epsilon_{ef0}$, and $X_{0}$ denote the values of the self-induction coefficient, radiative induction, time-averaged electromagnetic energy, time-averaged power radiation, effective induced electromotive force and leakage reactance in absence of the ME response, respectively. For seek of clarity, we have introduced the auxiliary functions $h(r_{1},r_{2})=(r_{1}^4+r_{2}^4-2r_{1}^3r_2)$, $p(r_{1},r_{2})=r_{1}^2(-r_{1}^4+10(r_{1}r_{2})^2)+9r_{2}^6-2(r_1r_2)^3(9+2\log(r_{1}/r_{2}))$ and $q(r_{1},r_{2})=r_{1}^6+2(r_{1}^2r_{2})^2-6(r_{1}r_{2})^3+3r_{2}^6$.}
\label{tab:table1}
\end{table*}

\subsection{Solenoid actuator}\label{Sec_FMHall}

Finally, we address the solenoid actuator in the magnetoquasitatic regime that consists of the solenoid previously treated, but now it is composed by a slideably disposed high-permeability cylindrical core that is partially inserted certain finite length $z_{0}$ at rest. Notice that compact actuators that flip latches or switches usual are usually implemented using solenoids \cite{apicella20191}.  From the conclusion drawn in Sec. \ref{Sec_BLS}, one could expect that the azimuthal surface Hall current appearing in the rod will induce an axial magnetization in its bulk, which could interact with the vacuum magnetic field generated by the exciting current flowing in the coil. According to standard electrodynamics, the latter will give rise to a magnetic force, say $F_{\text{Hall}}$, (which is given by Eq. (\ref{EqMF}) in App. \ref{app1}) that would pull the rod into the solenoid. Hence, it would be interesting to estimate the strength of $F_{\text{Hall}}$ up to first order in the ME susceptibility and exciting frequency. It can be shown (see App. \ref{app1}) that this can be obtained from the expression
\begin{align}
F_{\text{Hall}}
= &2\mu_{0}\alpha_{0}\int_{-(l-z_{0})}^{z_{0}} \int_{0}^{r_{2}} \theta(r,z)
\Big[
E_{z}(r,z)\frac{\partial H_{0,z}(r,z)}{\partial z} \nonumber \\
& \quad +
E_{r}(r,z)\frac{\partial H_{0,z}(r,z)}{\partial r}\Big]r drdz,
    \label{EFHall}
\end{align}
where $H_{0,z}$ is the axial magnetic field component in absence of the rod (their expressions are well-known \cite{lin20211}). 
As we are interested in a leading-order estimation of the force strength in the magnetoquasistatic regime, we have considered that both the fringe effects upon the magnetic rod and $H_{0,r}$ as well as $H_{0,\varphi}$ are negligible, which is true for thick and tall solenoids \cite{zangwill20121}. Under these prescriptions, on one hand, the second term within the integral in (\ref{EFHall}) can be further ignored, and on other hand, the axial component of the electric field can be approximated by Eq. (\ref{ELFAPZ_M}). After performing the integration once replaced the electric and magnetic fields, we get the following expression for the magnetic force for sufficiently low frequencies (i.e. $r_{2}\omega_{0}/c_{3} \ll 1$) and weak ME effects (i.e. $r_{2}\mu_{0}\omega_{0}\alpha_{0} \ll 1$), 
\begin{align}
F_{\text{Hall}}
&\approx\frac{r_{2}^{6}\mu_{0}^2(\mu_{2,r} \omega_{0}\alpha_{0} nI_{0})^2\theta_{2}(\theta_{2}-\theta_{3})\cos(\omega_{0}t)}{12l(\mu_{2,r}+1)} \nonumber \\
&  \times 
\Big[\frac{1}{z_{0}^2}
+\frac{1}{(z_{0}-2l)^2} -\frac{1}{(l-z_{0})^2}-\frac{1}{(l+z_{0})^2}\Big],
    \label{EFHall1}
\end{align}
where one could see that $F_{\text{Hall}}>0$ if $z_{0}<l/2$, whereas $F_{\text{Hall}}<0$ if $z_{0}>l/2$. Accordingly, Eq.(\ref{EFHall1}) reveals that the $F_{\text{Hall}}$ represents a second-order correction to the conventional magnetic force that pulls the magnetic rod into the solenoid. Hence, it can be concluded that TME materials does not render an advantage to implement actuators compared with multiferroic heterostructures. As similarly occurs to the self-induction coefficient, beyond the magnetoquasistatic regime one may expect that
a higher ME susceptibility could retrieve a larger magnetic force strength compared with the conventional situation. This question will be investigated elsewhere. 

\section{Outlook and concluding remarks}\label{SecCOUT}
In this paper we have presented a theoretical framework
to explore electromagnetic properties of passive magnetic devices composed of materials featuring an uniform ME response in the bulk; in particular, we have paid special attention to time-reversal-invariant TIs or AIs, which exhibit a ME susceptibility quantized in terms of the fine-structure constant. By following the magnetic circuit approach, we derived an extended version of the Hopkinson's law valid for both topological and non-topological ME materials in the magnetoquasistatic domain that takes account the azimuthal surface Hall current responsible for the ME effect, and also obtain the circuit diagram equivalent to the archetypal model of magnetic circuits, i.e. the (primitive) transformer. We discover that the benefit of employing TME materials is essentially twofold: the appearance of an additional perturbative source of magnetic flux regarding a second-order correction to the induction coefficients, and most importantly, the suppression of eddy currents since the topological core has virtually an insulating bulk. Although we have restricted our analysis to the regime where the magnetization is linear, the nonlinear magnetization effects could be approximately included in our results by previous experimental estimations of the hysteresis loop losses \cite{fusil20141,cherifi20141,liang20211,tian20171}.  Let us emphasize that our analysis is meaningful when strong absorption effects of the ME medium are negligible (that is, when its dielectric constant changes softly with frequency).

Our treatment also proves convenient to address electromagnetic properties of the ac solenoid inductor for both low and high frequencies as well as for weak and strong ME responses: we summarize in Table \ref{tab:table1} the results obtained for the self-induction coefficient, radiative coefficient, time-averaged electromagnetic energy, effective electromotive force, time-averaged  power radiated and magnetic linkage for three arrangement of the cylindrical TI layers. Our theory is particularly relevant for ME materials endowed with a relatively large ME susceptibilities (e.g. when it exceeds by a factor of $10^3$ the fine-structure constant), for which we estimate a self-inductance tunability of over $200\%$ up to $100$ GHz for both NI-TI and TI-TI configurations in the millimeter length scale (as the amplitude of the axial magnetic field significantly grows by the azimuthal surface Hall current). This result supports the idea that ME materials in the near future could represent a promising platform for the implementation of integrable tunable inductors in the RF domain \cite{philip20171,chen20201,he20211}.  

Additionally, our treatment predicts that the generation of highly-intense electromagnetic fields becomes energetically expensive in the presence of a strong ME response in comparison with the conventional situation, in turn the operating range would be restricted to frequencies sufficiently small for a given ME susceptibility: for example, the cut-off frequency in certain 2D multiferroic heterostructures, such as the  FeRh/BaTiO$_3$ compound, is degraded to the very low frequency domain in the millimeter length scale. Similarly, several ME inductors consisting of Metglas/PZT composites exhibit a monotonic decreasing of the inductance with the exciting frequency due to an eddy current screening effect \cite{su20161}. Here, it is important to recall that the dependence of the ME susceptibility with the exciting frequency has been neglected, which approximately holds in most of the studied scenarios \cite{fiebig20051}.

Remarkably, the recent developments on the fabrication
and manipulation of both topological and non-topological ME materials \cite{ying20221,fusil20141,nenno20201,bhattacharyya20211}
make them promising candidates to built passive magnetic electronic \cite{liang20211,gilbert20211} or low-power spintronics \cite{he20221,breunig20211,tian20171,tokura20191}, which open new avenues to implement
higher sophisticated technologies: ranging from power converters to RF communications. In particular
this prospect highlights the demand for further theoretical
tools to enable us to assess its feasible electromagnetic properties. In this context, the present treatment could render a valuable
theoretical guideline to design a new series of experiments in
the realm of basic ME technology.

\acknowledgments
The authors are grateful to A. Wagemakers for fruitful
discussions. This material is based upon work supported by the Nazarbayev University Faculty Development Competitive Research Grants Program 11022021FD2921. A. A. Valido acknowledge financial support from the
Spanish State Research Agency (AEI) and the European
Regional Development Fund (ERDF, EU) under project
PID2019-105554GB-I00.


\newpage

\appendix

\counterwithin{figure}{section}
\begin{widetext}
\section{Boundary conditions, the electromagnetic energy and the magnetic force}\label{app1}
In this appendix we briefly illustrate the derivation of boundary conditions (\ref{ABoundCond1})-(\ref{ABoundCond4}), the electromagnetic energy (\ref{EUEM}) and the magnetic force (\ref{EFHall}) from the $\theta$-electrodynamics. We start from the $\theta$-electrodynamics Lagrangian density, this can be expressed as follows \cite{essin20091,qi20081,qi20111,nomura20111,nenno20201,sekine20211}:
\begin{equation}
   \mathcal{L}=\frac{1}{2}\bigg(\varepsilon\vect E^2-\frac{1}{\mu}\vect B^2\bigg)+\frac{\alpha_{0}\theta }{\pi}\vect E\cdot \vect B,  \label{ELDAE}
\end{equation}
recall that $\varepsilon=\varepsilon_{r}\varepsilon_{0}$ and $\mu=\mu_{r}\mu_{0}$. The first term in the right-hand side of (\ref{ELDAE}) corresponds to the well-known Maxwell kinetic term, and the second term is responsible for the ME response. This can be readily seen by recalling that $-\vect D$ represents the canonical momentum conjugate to the vector potential $\vect A$ and $\vect H$ is the canonical conjugate to $\vect B$ \cite{nogueira20221}, Eq. (\ref{ELDAE}) thus yields 
\begin{align*}
    \vect D&=\frac{\partial \mathcal{L}}{\partial (-\dot{\vect A})}=\frac{\partial \mathcal{L}}{\partial \vect E}=\varepsilon\vect E+\frac{\alpha_{0}\theta}{\pi}\vect B , \nonumber \\
    \vect H&= -\frac{\partial \mathcal{L}}{\partial \vect B}=\frac{1}{\mu}\vect B-\frac{\alpha_{0}\theta}{\pi}\vect E,
\end{align*}
which coincides with the constitutive relations (\ref{ECT2}) and (\ref{ECT1}) characteristic of the ME response. Now the electromagnetic Hamiltonian is obtained from (\ref{ELDAE}) by doing the usual Legendre's transform, that is
\begin{align*}
    \mathcal{H}&=-\dot{\vect A}\cdot \vect D-\mathcal{L}, \nonumber  \\
     &=\frac{1}{\varepsilon}\Big(\vect D-\frac{\alpha_{0}\theta}{\pi}\vect B\Big)\cdot \vect D-\frac{1}{2 \varepsilon}\Big(\vect D-\frac{\alpha_{0}\theta}{\pi}\vect B\Big)^2+\frac{1}{2 \mu}\vect B^2-\frac{\alpha_{0}\theta}{\pi\varepsilon}(\vect D-\frac{\alpha_{0}\theta}{\pi}\vect B)\cdot \vect B \nonumber \\
     &=\frac{1}{\varepsilon}\vect D^2 -\frac{\alpha_{0}\theta}{\pi\varepsilon}\vect B \cdot \vect D-\frac{1}{2 \varepsilon}\Big(\vect D^2+\Big(\frac{\alpha_{0}\theta}{\pi}\Big)^2\vect B^2-2\frac{\alpha_{0}\theta}{\pi}\vect B\cdot \vect D\Big)+\frac{1}{2 \mu}\vect B^2-\frac{\alpha_{0}\theta}{\pi \varepsilon}\Big( \vect D-\frac{\alpha_{0}\theta}{\pi}\vect B\Big)\cdot \vect B \nonumber \\
     &=\frac{1}{2 \varepsilon}\vect D^2+\frac{1}{2 \mu}\Bigg(1+\frac{\mu}{\varepsilon}\bigg(\frac{\alpha_{0}\theta}{\pi }\bigg)^2\Bigg)\vect B^2-\frac{\alpha_{0}\theta}{\pi\varepsilon}\vect B\cdot \vect D. 
\end{align*}
After replacing $\vect D=\varepsilon\vect E+\alpha_{0}\theta/\pi\vect B$ in term of the fundamental fields and integrating over the system volume $\mathcal{V}$, we recover the expression (\ref{EUEM}) for the electromagnetic energy, as desired. One can also obtain the modified Maxwell's equations from the Lagrange's equations associated to (\ref{ELDAE}).

As we are interested in bilayer solenoid inductors that present a small discontinuity across their cylindrical surfaces, generically denoted by $\Sigma$, it is important to pay attention to the boundary conditions in order to solve the electromagnetic fields. Given the modified Maxwell's equations (\ref{ME1})-(\ref{ME4}), these can be expressed as follows
\begin{align}
    \vect n\times\big[\vect E\big]_{\Sigma}=&0, \label{EBoundCond1} \\
    \vect n\cdot\big[\vect B\big]_{\Sigma}=&0, \label{EBoundCond2} \\
    \vect n\cdot\big[\varepsilon\vect E\big]_{\Sigma}=&\sigma_{f}-\frac{\alpha_{0}}{\pi}\vect n\cdot[\theta\vect B]_{\Sigma}, \label{EBoundCond3} \\
    \vect n\times\bigg[\frac{\vect B}{\mu}\bigg]_{\Sigma}=&\vect K_{f}+\frac{\alpha_{0}}{\pi}\vect n\times[\theta\vect E]_{\Sigma}, \label{EBoundCond4} 
\end{align}
where $\vect n$ is the outer unitary normal vector to the surface $\Sigma$, and  $\big[\vect Z\big]_{\Sigma}=\vect Z(\Sigma^{+})-\vect Z(\Sigma^{-})$. Furthermore, $\vect K_{f}$ and $\sigma_{f}$ are, respectively, the free current and charge densities on the boundary surface $\Sigma$. By replacing the electric and magnetic fields in terms of the vector potential (i.e. $\vect E=-\nabla \vect \phi-\partial_{t}\vect A$ and $\vect B=\nabla \times \vect A$), one can show that these expressions returns Eqs. from (\ref{ABoundCond1}) to (\ref{ABoundCond4}). 

Finally, we would like to mention the magnetic force (\ref{EFHall}) arising in the solenoid actuator. From standard electrodynamics it is well-known that a magnetization $\vect M$ of an isolated material in presence of an external magnetic field $\vect H_{0}$ gives rise to a magnetic force given by the expression \cite{zangwill20121,zahn19791,jackson20121}:
\begin{align*}
\vect F=\mu_{0}\int_{\mathcal{V}} (\vect M\cdot \nabla)\vect H_{0}\ dr^3.
\end{align*}
By paying attention to the constitutive relation (\ref{ECT1}) one may identify the magnetization vector, which yields
\begin{align}
\vect F=\mu_{0}\int_{\mathcal{V}}(\vect M_{0}(\vect r)\cdot \nabla)\vect H_{0}(\vect r) \ dr^3
   -\frac{\mu_{0}\alpha}{\pi}\int_{\mathcal{V}} \theta(\vect r)(\vect E(\vect r)\cdot\nabla)\vect H_{0}(\vect r) \ dr^3,\label{EqMF}
\end{align}
where $\vect M_{0}$ denotes the magnetization vector in absence of the ME effect, i.e. $\vect M_{0}(\vect r)=(\mu(\vect r)/\mu_{0}-1)\vect H_{0}(\vect r)$ for linear magnetic media. Hence, the second term on the right-hand side of Eq. (\ref{EqMF}) represents the magnetic force exclusively emerging from the ME response, which has been called $F_{\text{Hall}}$. This result is used in Sec.\ref{Sec_FMHall} to obtain an approximate expression of the magnetic force due to the surface Hall currents (see Eq. (\ref{EFHall1})).

\section{Solutions of the bilayer long solenoid}\label{AppBLSSol}

Here we briefly illustrate the procedure to obtain the solutions of Eqs. (\ref{EoMAP})-(\ref{EoMAZ}) together with boundary conditions (\ref{ABoundCond1})-(\ref{ABoundCond4}). As stated in Sec. \ref{Sec_BLS}, it is convenient to express these solutions as linear combinations of the Bessel functions of first kind  $J_{1}(kr)$ and third kind $H_{1}^{(1)}(kr)$ (see Ref.\cite[pp. 112-116]{jackson20121}, Ref.\cite[pp. 718]{zangwill20121},  Ref.\cite[pp. 162]{chew19951} and Ref.\cite{harmon19911}) as shown in (\ref{ESAPhi}) and (\ref{ESAz}), where we have to determine the auxiliary coefficients: $a_{\varphi},a_{z},b_{\varphi},b_{z},c_{\varphi},c_{z},d_{\varphi},d_{z}$. Notice that we have made use of the fact that $H_{1}$ diverges at $r\rightarrow 0$. After imposing the boundary conditions, we obtain the following linear system of equations from which elucidate the aforementioned coefficients,
\begin{equation}
    \vect L\cdot(a_{\varphi},a_{z},b_{\varphi},b_{z},c_{\varphi},c_{z},d_{\varphi},d_{z})^{T}=(0,0,0,0,0,0,-4\pi n I_{0},0)^{T}
    \label{LASE}
\end{equation}
with $\vect L$ given by,
\begin{equation}
\left( \begin{array}{cccccccc}
J_{1}(k_{1}r_{1}) & 0  & -J_{1}(k_{2}r_{1}) & 0 & H_{1}^{(1)}(k_{2}r_{1}) & 0 & 0 & 0  \\
0 &J_{1}(k_{1}r_{1}) & 0  & J_{1}(k_{2}r_{1}) & 0 &  -H_{1}^{(1)}(k_{2}r_{1}) & 0  & 0 \\
0 & 0  &  J_{1}(k_{2}r_{2}) & 0 &  H_{1}^{(1)}(k_{2}r_{2})  & 0  & - H_{1}^{(1)}(k_{3}r_{2})  & 0 \\
0 &0 & 0  & J_{1}(k_{2}r_{2}) & 0 &  H_{1}^{(1)}(k_{2}r_{2})  &0  &  -H_{1}^{(1)}(k_{3}r_{2})  \\
\frac{k_{1}}{\mu_{1}}J_{0}(k_{1}r_{1}) & -a_{\theta}J_{1}(k_{1}r_{1})   &\frac{k_{1}}{\mu_{2}}J_{0}(k_{2}r_{1}) & 0  & \frac{k_{2}}{\mu_{2}}H_{0}^{(1)}(k_{2}r_{1}) & 0 & 0 & 0   \\
a_{\theta}J_{1}(K_1r_1) &  -\frac{k_1}{2\mu_1}f(1,1) &0& \frac{k_2}{2\mu_2}f(2,1) & 0  & \frac{k_2}{2\mu_2}g(2,1)  &  0  &0\\
0 & 0& -\frac{k_{2}}{\mu_{2}}J_{0}(k_{2}r_{2}) &- b_{\theta}J_{1}(k_{2}r_{2}) & -\frac{k_{2}}{\mu_{2}}H_{0}^{(1)}(k_{2}r_{2})  & -b_{\theta}H_{1}^{(1)}(k_{2}r_{2}) & \frac{k_{3}}{\mu_{3}}H_{0}^{(1)}(k_{3}r_{2}) & 0  \\
0  & 0 &ib_{\theta}J_{1}(k_2r_2) & -\frac{k_2}{2\mu_2}f(2,2)  & ib_{\theta}H_{1}^{(1)}(k_2r_2) & -\frac{k_2}{2\mu_2}g(2,2) & 0 & \frac{k_3}{2\mu_3}g(3,2)  
\end{array}\right) ,
\label{DLQUMI}
\end{equation}
where we have introduced the auxiliary elements
\begin{align*}
    a_{\theta}&= \frac{i\omega_{0}\alpha_{0}}{\pi}(\theta_{2}-\theta_{1}), \nonumber \\
    b_{\theta}&= \frac{i\omega_{0}\alpha_{0}}{\pi}(\theta_{3}-\theta_{2}), \nonumber \\
    f(i,j)&= J_{0}(k_ir_j)-J_{2}(k_ir_j),   \nonumber \\
    g(i,j)&=H^{(1)}_{0}(k_ir_j)-H^{(1)}_{2}(k_ir_j).
\end{align*}
In particular, 
the last row in (\ref{DLQUMI}) is the so-called jump condition (see Ref.\cite[pp. 256]{zangwill20121} for further details). We make use of the symbolic computation handled by 
MATHEMATICA to solve (\ref{LASE}) and completely determine the coefficients $a_{\varphi},a_{z},b_{\varphi},b_{z},c_{\varphi},c_{z},d_{\varphi},d_{z}$. As explained in Secs. \ref{SecLFR} and \ref{SecHFR}, these solutions can be approximated in the low and high frequency regime as well as for small and large ME susceptibilities, by combining the Taylor's expansion with the well-known asymptotic forms of the Bessel functions.

\subsection{Approximate solutions of electromagnetic fields for non-permeable media}

In this section we illustrate the electromagnetic fields in the domain of small ME susceptibility and low frequency, i.e. $r_{2}\omega_{0}/c_{i}\ll 1$ (for $i=1,2,3$) and $r_{2}\mu_{0}\omega_{0}\alpha_{0}\ll 1$. Concretely, we obtain the following expressions for the magnetic field for $r_{1}<r<r_{2}$,
\begin{align}
&B_{\varphi}(r_{1}<r<r_{2},t)
\approx  
\mu_{0}^2\frac{n I_{0}}{4\pi l}\frac{\alpha_{0}\omega_{0}}{r^2}
\Big[
r_{1}^3(\theta_2 - \theta_1) +r^2r_{2}(\theta_2 - \theta_3)
\Big]\sin(\omega_{0}t), \label{BLFAPZ2a} \\
&B_{z}(r_{1}<r<r_{2},t)
\approx
\pi\mu_{0}\frac{n I_{0}}{4l}\Big(\frac{r_{2}\omega_{0}}{c_{3}}\Big)^2\sin(\omega_{0}t)
\label{BLFAPZ2b} \\
& 
+
\mu_{0}\frac{n I_{0}}{4\pi l}
\Big[
4\pi+\frac{(\mu_{0}\alpha_{0}\omega_{0})^2}{\pi r_{2}}
\Big(r_{2}^3(\theta_{2}-\theta_{3})+r_{1}^3(\theta_1-\theta_2))(\theta_{2}-\theta_{3})
\Big)  +
\pi\Big(\frac{r_{2}\omega_{0}}{c_{3}}\Big)^2 
\Big\{ 1 +2\Big(\log\Big(\frac{2c_{3}}{r_{2}\omega_{0}}\Big)-\gamma_{Euler}\Big)\Big\}\Big]\cos(\omega_0 t), \nonumber
\end{align}
and the electric field,
\begin{align}
E_{\varphi}(r_{1}<r<r_{2},t)
&\approx 
\mu_{0}\frac{n I_{0}}{4\pi l} \omega_{0}\Big[2\pi r 
+
\frac{(\mu_{0}\alpha_{0}\omega_{0})^2}{2\pi r r_{2}}
\Big(r_{1}^4r_{2}(\theta_{1}-\theta_2)^2+r_{1}^3(r^2+r_{2}^2)(\theta_{1}-\theta_2)(\theta_{2}-\theta_{3})+r^2r_{2}^3(\theta_{2}-\theta_{3})^2
\Big)\nonumber \\
& \qquad \qquad \quad +
\pi \omega_{0}^2\Big(\frac{rr_{2}^2}{2c_{2}^2}+\frac{r_{1}^4}{2r}\Big(\frac{1}{c_{1}^2} -\frac{1}{c_{2}^2}\Big)\Big)+\pi r\Big(\frac{r_{2}\omega_{0}}{c_{3}}\Big)^2\Big(\log\Big(\frac{2c_{3}}{r_{2}\omega_{0}}\Big)-\gamma_{Euler}\Big)\Big]\sin(\omega_{0}t) \nonumber \\
&\quad -
\pi\mu_{0}\omega_{0}r\frac{n I_{0}}{8l}\Big(\frac{r_{2}\omega_{0}}{c_{3}}\Big)^2\cos(\omega_{0}t), \label{ELFAPa} \\
E_{z}(r_{1}<r<r_{2},t)
&\approx 
\mu_{0}^2\frac{nI_{0}}{4\pi l}\frac{\alpha_{0}\omega_{0}^2}{r}
\Big[
r_{1}^3(\theta_1 - \theta_2) +r^2r_{2}(\theta_2 - \theta_3)
\Big]
\cos(\omega_{0}t).
    \label{ELFAPb}
\end{align}
Similarly, we find out that the approximate magnetic field outside the solenoid inductor takes the form
\begin{align}
B_{\varphi}(r_{2}<r,t)
&\approx 
-\frac{n I_{0}}{16\pi l}\frac{ \pi\mu_{0}^2\omega_{0}^3\alpha_{0}}{ c_{3}^2}
\Big(
r_{1}^3(\theta_1 - \theta_2) +r_{2}^3(\theta_2 - \theta_3)
\Big) \nonumber \\
&\qquad \times
\Big[
\Big(J_{0}(k_{3}r)-J_{2}(k_{3}r)\Big)
\cos(\omega_{0}t)
+
\Big(N_{0}(k_{3}r)-N_{2}(k_{3}r)\Big)
\sin(\omega_{0}t)
\Big],  \label{BLFAPa} \\
B_{z}(r_{2}<r,t)
&\approx 
\frac{n I_{0}}{16\pi l}\frac{\mu_{0}\omega_{0}^2}{c_{2}^2}
\Big[
(2\pi r_{2})^2
+
(\mu_{0}\alpha_{0}\omega_{0})^2
\Big(r_{2}^4(\theta_{2}-\theta_{3})^2+2r_{1}^3r_{2}(\theta_1-\theta_2)(\theta_{2}-\theta_{3})+r_{1}^4(\theta_1-\theta_2)^2\Big) \nonumber \\
& \qquad + 
(\pi\omega_{0})^2
\Big\{r_{1}^4\Big(\frac{1}{c_{1}^2}-\frac{1}{c_{2}^2}\Big)+r_{2}^4\Big(\frac{1}{c_{2}^2}-\frac{1}{c_{3}^2}\Big)\Big\}
\Big]
\Big(J_{0}(k_{3}r)\sin(\omega_{0}t) -N_{0}(k_{3}r)\cos(\omega_{0}t)\Big),
    \label{BLFAPb}
\end{align}
and the electric field,
\begin{align}
E_{\varphi}(r_{2}<r,t)
&\approx 
-\frac{n I_{0}}{16\pi l}\frac{\mu_{0}\omega_{0}^2}{c_{3}}
\Big[
(2\pi r_{2})^2
+
(\mu_{0}\alpha_{0}\omega_{0})^2
\Big(r_{2}^4(\theta_{2}-\theta_{3})^2+2r_{1}^3r_{2}(\theta_1-\theta_2)(\theta_{2}-\theta_{3})+r_{1}^4(\theta_1-\theta_2)^2
\Big) \nonumber \\
& \qquad \qquad \qquad \quad +
(\pi\omega_{0})^2
\Big\{
r_{1}^4\Big(\frac{1}{c_{1}^2}-\frac{1}{c_{2}^2}\Big)+r_{2}^4\Big(\frac{1}{c_{2}^2}-\frac{1}{c_{3}^2}\Big)
\Big\}\Big] \nonumber \\
& \qquad \qquad \qquad  \times
\Big[
\Big(J_{0}(k_{3}r)-J_{2}(k_{3}r)\Big)\cos(\omega_{0}t)
+
\Big(N_{0}(k_{3}r)-N_{2}(k_{3}r)\Big)\sin(\omega_{0}t)
\Big], \label{ELFAPbou}\\   
E_{z}(r_{2}<r,t)
&\approx 
\pi\frac{n I_{0}}{8\pi l}\frac{\mu_{0}^2\alpha_{0}\omega_{0}^3}{c_{3}}
\big[r_{1}^3(\theta_1 - \theta_2) +r_{2}^3(\theta_2 - \theta_3)\big] 
\Big[J_{0}(k_{3}r)\sin(\omega_{0}t) -N_{0}(k_{3}r)\cos(\omega_{0}t)\Big].
    \label{ELFAZbou}
\end{align}
Equations from (\ref{BLFAPZ2a}) to (\ref{ELFAZbou}) were used to compute the induction coefficients  (\ref{EqSelfindLowT}) and (\ref{NradCoef}) as well as  the time-averaged electromagnetic energy (\ref{EEMELS}). Expressions from (\ref{BLFAPa}) to (\ref{ELFAPbou}) return (\ref{PradCoef}) the time-averaged radiative power after replaced in (\ref{PWRC}). Let us emphasize that the expression for the light speeds $c_{i}$ (with i=1,2) should be replaced by Eq. (\ref{EQDISLF}) according to the discussion of Sec. \ref{SecLFR}.

We also compute approximate expressions for the electromagnetic fields when dealing the TI-TI solenoid inductor (recall $r_{1}=r_{2}$ and $\theta_{1}=\theta_{2}$) for high frequencies and strong ME effects. The magnetic field components take the form in the inner region:
\begin{align}
B_{\varphi}(r<r_{2},t)
&\approx  
\frac{\sqrt{r_{2}\pi^5k_2^3}nI_{0}
\Big(J_{0}(k_2 r)-J_{2}(k_2 r)\Big)\sin(\omega_{0}t)}{2\pi l(\alpha_{0}\omega_{0})(\theta_{2}-\theta_{3})\Big(\cos(k_2r_{2})-\sin(k_2r_{2})\Big)}, \label{BLFAPPSTITI} \\
B_{z}(r<r_{2},t)
&\approx  
-\frac{\sqrt{2r_{2}\pi^7k_2^3}nI_{0}\sec(\pi/4+k_2r_2)J_{0}(k_2 r)
\Big(k_{3}\sin(\omega_{0}t)+k_2\tan(\pi/4+k_{2}r_{2})\cos(\omega_{0}t)\Big)}{2\pi l\mu_{0}(\alpha_{0}\omega_{0})^2\left(\theta_{2}-\theta_{3}\right)^2},   \label{BLFAPZSTITI}
\end{align}
and the electric field components,
\begin{align}
E_{\varphi}(r<r_{2},t)
&\approx  
\frac{\sqrt{2r_{2}\pi^7k_2^3}nI_{0}\sec(\pi/4+k_2r_2)J_{1}(k_2 r)
\Big(k_{3}\cos(\omega_{0}t)-k_2\tan(\pi/4+k_{2}r_{2})\sin(\omega_{0}t)\Big)}{2\pi l\mu_{0}\alpha_{0}^2\omega_{0}\left(\theta_{2}-\theta_{3}\right)^2}, \label{ELFAPPSTITI} \\
E_{z}(r<r_{2},t)
&\approx 
\frac{\sqrt{r_{2}\pi^5k_2}nI_{0}J_{1}(k_2 r)\cos(\omega_{0}t)}{2\pi l\alpha_{0}(\theta_{2}-\theta_{3})\Big(\cos(k_2r_{2})-\sin(k_2r_{2})\Big)},\label{ELFAP1ZSTITI}
\end{align}
where we must take account the dispersion relation (\ref{EQDISHF}) according to the discussion in Sec. \ref{SecHFR}. From Eqs. (\ref{BLFAPPSTITI}), (\ref{BLFAPZSTITI}), (\ref{ELFAPPSTITI}) and (\ref{ELFAP1ZSTITI}) we obtain the induction coefficients (\ref{LSelfInMEH}) and (\ref{NSelfInMEH}) and the electromagnetic energy (\ref{UMEH}) in the high frequency and strong ME susceptibility domain. 

By making use of Eqs. (\ref{BLFAPZ}) and (\ref{BLFAPZ2b}), we have also obtained a reduced expression for the effective electromotive force at leading order for weak ME effects and low frequencies,
\begin{align}
    \epsilon_{\text{ef}}&\approx \pi r_2^2\mu_{0}\frac{n I_{0}}{2\sqrt{2}l}\omega_{0}\Bigg(2+\frac{1}{2}\Big(\frac{r_2\omega_0}{c_{3}}\Big)^2\Big[\Big(\frac{c_{3}r_{1}^2}{ c_{1}}\Big)^2 +\Big(\frac{c_{3}}{c_2}\Big)^2(r_{2}^4-r_{1}^4)+2r_{2}^4\Big(\log\Big(\frac{2c_{3}}{r_{2}\omega_{0}}\Big)-\gamma_{Euler}\Big)\Big]  \nonumber \\
    &+\frac{(\mu_{0}\alpha_{0}\omega_{0})^2}{2\pi^2r_{2}^2}\big[r_{1}^4\left(\theta_{1}-\theta_{2}\right)^2 +2r_1^3r_2\left(\theta_{1}-\theta_{2}\right)\left(\theta_{2}-\theta_{3}\right) +r_{2}^4\left(\theta_{2}-\theta_{3}\right)^2 \big]\Bigg), 
    \label{EQEMTFV}
\end{align}
 where the light speed in the ME medium is approximated by Eq. (\ref{EQDISLF}), as explained in Sec. \ref{SecLFR}. Additionally, we provide Fig. (\ref{FigErMEC}) which illustrates  electromagnetic fields as a function of the radial distance and the ME susceptibility. These results are discussed in Sec. \ref{Sec_NCEMF}.

\begin{figure}
\centering
\includegraphics[scale=0.32]{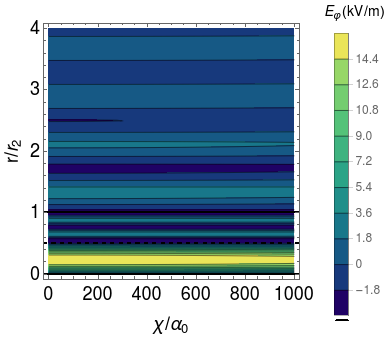}
\includegraphics[scale=0.32]{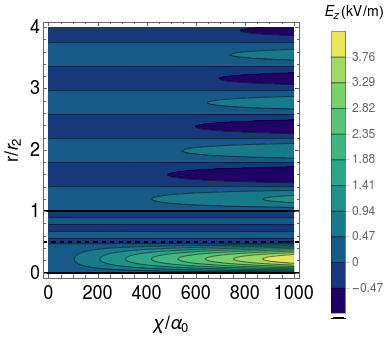}
\includegraphics[scale=0.32]{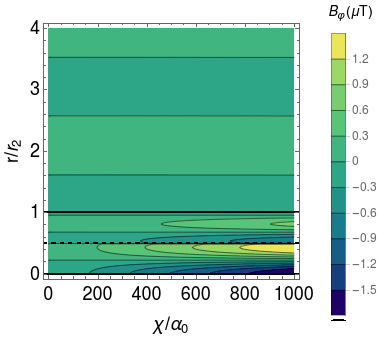}
\includegraphics[scale=0.32]{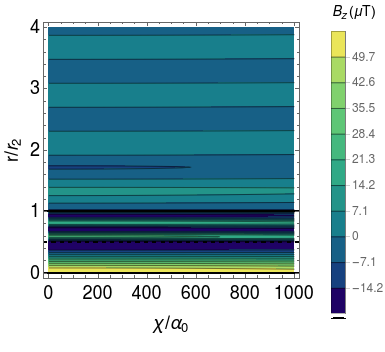}
\includegraphics[scale=0.32]{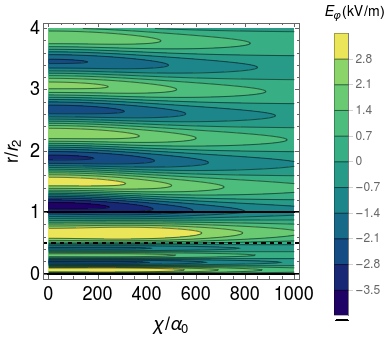}
\includegraphics[scale=0.32]{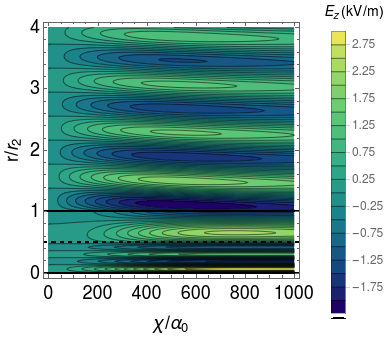}
\includegraphics[scale=0.32]{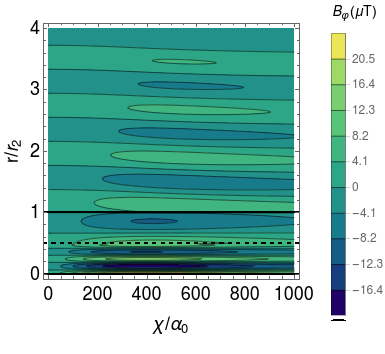}
\includegraphics[scale=0.32]{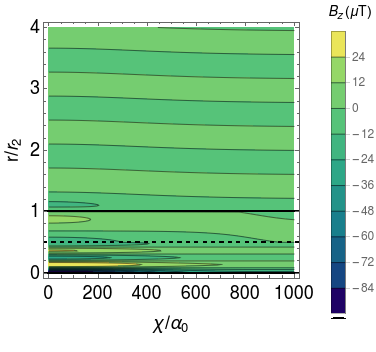}
\includegraphics[scale=0.32]{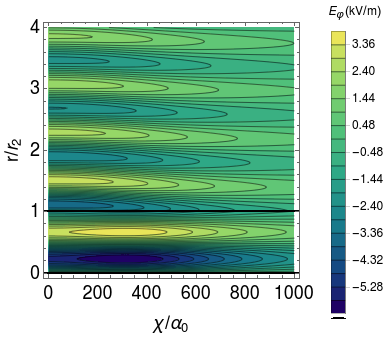}
\includegraphics[scale=0.32]{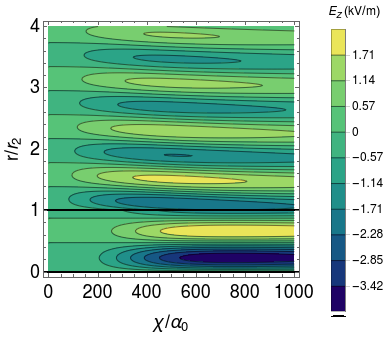}
\includegraphics[scale=0.32]{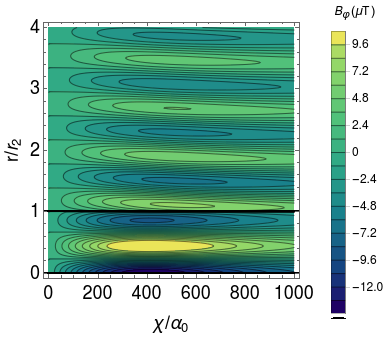}
\includegraphics[scale=0.32]{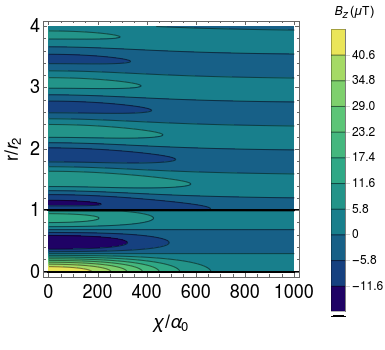}
\caption{(color online). Contour plots of electromagnetic fields as a function of the radial length and the ME susceptibility for the three solenoid inductors sketched in Fig. (\ref{Fig1Setup}): TI-NI (upper), NI-TI (central), TI-TI (down). In all pictures, we consider $nI_{0}/l=1$ A/m, $\omega_{0}=24.1$ THz, and $t=\pi/\omega_{0}$, whereas the solenoid inner and outer radius were taken $r_{1}=0.05$ mm and $r_{2}=0.1$ mm, respectively. We consider non-permeable media $\mu_{1,r}=\mu_{2,r}=1$.  \label{FigErMEC}}
\end{figure} 

\subsection{Approximate solutions of electromagnetic fields for permeable media}

Now we illustrate the simplified expressions of electromagnetic fields for a permeable medium in the case of the TI-TI solenoid inductor. We first study the low frequency, weak ME susceptibility and high permeability domain: that is, $r_{2}\mu_{0}\omega_{0}\alpha_{0} \ll 1$, $r_{2}\omega_{0}/c_{3} \ll 1$ and $ \mu_{2,r}\gg 1$. Upon carrying a perturbative analysis, we obtain the following expression for the magnetic field,
\begin{align}
B_{\varphi}(r<r_{2},t)
&\approx  
\frac{\mu_{0}^2\mu_{2,r}^2n I_{0}\alpha_{0}\omega_{0}
r_{2}(\theta_2 - \theta_3)}{2\pi l(\mu_{2,r}+1)}
\sin(\omega_{0}t),\label{BLFAPP_M} \\
B_{z}(r<r_{2},t)
&\approx  \frac{n I_{0}\mu_{2,r}}{4 \pi^2 l }
\Big[
4\pi^2\mu_{0}+\frac{2(\mu_{0}\mu_{2,r} r_{2}\alpha_{0}\omega_{0}(\theta_{2}-\theta_{3}))^2}{(\mu_{2,r}+1)} \nonumber \\
& \qquad \qquad 
+\pi^2 \mu_{0}\Big(\frac{r_{2}\omega_{0}}{c_{2}}\Big)^2+2\pi^2\Big(\frac{r_{2}\omega_{0}}{c_{3}}\Big)^2\Big(\log\Big(\frac{2c_{3}}{r_{2}\omega_{0}}\Big)-\gamma_{Euler}\Big)\Big]
\cos(\omega_0 t) \nonumber \\
& \quad +\frac{\pi n I_{0}\mu_{0}\mu_{2,r}^2}{4l}\Big(\frac{r_{2}\omega_{0}}{c_{3}}\Big)^2\sin(\omega_{0}t),
    \label{BLFAPZ_M}
\end{align}
and the electric field,
\begin{align}
E_{\varphi}(r<r_{2},t)
&\approx 
\frac{n I_{0}\mu_{2,r} r}{4 \pi^2 l }
\Big[
4\pi^2\mu_{0}+\frac{2(\mu_{0}\mu_{2,r} r_{2}\alpha_{0}\omega_{0}(\theta_{2}-\theta_{3}))^2}{(\mu_{2,r}+1)} \nonumber \\
& \qquad \qquad \qquad +
\pi^2\mu_{0}\Big(\frac{\omega_{0}r_{1}^2}{c_{2}^2}\Big)+\pi \Big(\frac{r_{2}\omega_{0}}{c_{3}}\Big)^2\Big(\log\Big(\frac{2c_{3}}{r_{2}\omega_{0}}\Big)-\gamma_{Euler}\Big)\Big]\sin(\omega_{0}t) \nonumber \\
& \quad -
\frac{\pi nI_{0}\mu_{0}\mu_{2,r}^2\omega_{0} r}{8l}\Big(\frac{r_{2}\omega_{0}}{c_{3}}\Big)^2\cos(\omega_{0}t),  \label{ELFAPP_M} \\
E_{z}(r<r_{2},t)
&\approx \frac{n I_{0}\mu_{0}^2\mu_{2,r}^2 r_{2} r\alpha_{0}\omega_{0}^2(\theta_{2}-\theta_{3})}{2\pi l(\mu_{2,r}+1)}
\cos(\omega_{0}t),
    \label{ELFAPZ_M}
\end{align}
where $c_{2}$ must be replaced by Eq. (\ref{EQDISLF}). By following a similar procedure as illustrated in Sec.  \ref{SecLFR}, one can show that Eq. (\ref{BLFAPZ_M}) satisfies the expression (\ref{MEQS4}), which shows the validity of the modified Ampère's law in a permeable medium. Furthermore, Eq. (\ref{EEMTFLSME1}) in Sec. \ref{Sec_MCC} is obtained from (\ref{EEEMTF}) after some manipulation once replaced the magnetic flux retrieved by (\ref{BLFAPZ_M}).

\end{widetext}

\newpage
\bibliographystyle{apsrev4-1}
\bibliography{references1}

\end{document}